\documentclass[apj,twocolumn]{openjournal}
\usepackage{amsmath}

\usepackage[dvipsnames]{xcolor} 
\usepackage[breaklinks,colorlinks,citecolor=blue,urlcolor=blue]{hyperref}

\newcommand{\sfive}{\textit{S}$^5$}

\usepackage{orcidlink}

\renewcommand{\arraystretch}{1.2}  
\usepackage{listings}
\usepackage{color}
\definecolor{dkgreen}{rgb}{0,0.6,0}
\definecolor{gray}{rgb}{0.5,0.5,0.5}
\definecolor{mauve}{rgb}{0.58,0,0.82}
\definecolor{golden}{rgb}{0.86,0.65,0.01}
\usepackage{color}
\usepackage{soul}
\usepackage{amsmath}
\usepackage{xspace}
\usepackage{graphicx}
\usepackage{enumitem}
\usepackage{amssymb}
\usepackage{xifthen}
\usepackage{hyperref}
\usepackage[normalem]{ulem}
\usepackage{multirow}
\usepackage{lineno}
\usepackage{changepage}

\newcommand{\code}[1]{\texttt{#1}\xspace}

\newcommand{\gaia}{\textit{Gaia}\xspace}

\newcommand{\unit}[1]{\ensuremath{\mathrm{\,#1}}\xspace}
\newcommand{\feh}{\unit{[Fe/H]}}
\newcommand{\teff}{\ensuremath{T_\mathrm{eff}}\xspace}
\newcommand{\logg}{\ensuremath{\log\,g}\xspace}

\newcommand{\vhel}         {\mbox{$v_{\mathrm{hel}}$}}

\newcommand{\vt}         {\mbox{$\nu_{\mathrm{t}}$}}
\newcommand{\kms}{\unit{km\,s^{-1}}}
\newcommand{\masyr}{\unit{mas\,yr^{-1}}}
\newcommand{\msun}{\unit{M_\odot}}

\begin{document}


\title{Chemical Abundances in the Metal-Poor Globular Cluster ESO~280-SC06: A Formerly Massive, Tidally Disrupted Globular Cluster}

\author{Sam A. Usman\,\orcidlink{0000-0003-0918-7185}$^{1,2}$}
\author{Alexander P. Ji\,\orcidlink{0000-0002-4863-8842}$^{1,2,3}$}
\author{Jandrie Rodriguez\,\orcidlink{0009-0003-9490-3015}$^{4}$}
\author{Jeffrey~D. Simpson\,\orcidlink{0000-0002-8165-2507}$^{5,6}$}
\author{Sarah~L.~Martell\,\orcidlink{0000-0002-3430-4163}$^{5,6}$}
\author{Ting~S. Li\,\orcidlink{0000-0002-9110-6163}$^{7}$}
\author{Ana Bonaca\,\orcidlink{0000-0002-7846-9787}$^8$}
\author{Shivani P. Shah\,\orcidlink{0000-0002-3367-2394}$^9$}
\author{Madeleine McKenzie\,\orcidlink{0000-0002-1715-1257}$^8$}

\affiliation{$^1$Department of Astronomy \& Astrophysics, University of Chicago, 5640 S Ellis Avenue, Chicago, IL 60637, USA}
\affiliation{$^2$Kavli Institute for Cosmological Physics, University of Chicago, Chicago, IL 60637, USA}
\affiliation{$^3$NSF-Simons AI Institute for the Sky (SkAI), 172 E. Chestnut St., Chicago, IL 60611, USA}
\affiliation{$^4$Department of Physics, Syracuse University, Crouse Dr, Syracuse, NY 13210}
\affiliation{$^5$School of Physics, University of New South Wales, Sydney, NSW 2052, Australia}
\affiliation{$^6$Centre of Excellence for All-Sky Astrophysics in Three Dimensions (ASTRO 3D), Australia}
\affiliation{$^7$David A. Dunlap Department of Astronomy \& Astrophysics, University of Toronto, 50 St George Street, Toronto ON M5S 3H4, Canada}
\affiliation{$^8$Observatories of the Carnegie Institution for Science, 813 Santa Barbara St., Pasadena, CA 91101, USA}
\affiliation{$^9$Department of Astronomy, University of Florida, 211 Bryant Space Science Center, Gainesville, FL 32601, USA}
\email{Corresponding author: susman@uchicago.edu}

\begin{abstract}
We present the first high-resolution abundance study of ESO~280-SC06, one of the least luminous and most metal-poor gravitationally bound Milky Way globular clusters.
Using Magellan/MIKE spectroscopy for ten stars, we confirm the cluster's low metallicity as [Fe/H] = $-2.54 \pm 0.06$ and the presence of a nitrogen-enhanced star enriched by binary mass transfer.
We determine abundances or abundance upper limits for 21 additional elements from the light, alpha, odd-Z, iron peak, and neutron-capture groups for all ten stars. 
We find no spread in neutron-capture elements, unlike previous trends identified in some metal-poor globular clusters such as M15 and M92.
Eight of the ten stars have light-element abundance patterns consistent with second-population globular cluster stars, which is a significantly larger second-population fraction than would be expected from the low present-day mass of $10^{4.1}$ M$_{\odot}$. 
We estimate the initial mass of the cluster as $10^{5.4 – 5.7}$ M$_{\odot}$ based on its orbit in the Milky Way.
A preferential loss of first-population stars could explain the high fraction of second-population stars at the present time.
Our results emphasize the importance of considering mass loss when studying globular clusters and their enrichment patterns.
\end{abstract}

\maketitle
\section{Introduction} \label{sec:intro}
Globular clusters (GCs) were once thought to each be comprised of a simple stellar population where all stars had the same age and chemical abundance.
However, chemical analyses starting in the 1970s have identified unexpected, correlated abundance patterns within these clusters \citep{Osborn1971}.
Generally in GCs, we find stars with light element abundances similar to solar abundances when scaled to the correct metallicity (the \textit{first population}, or \textit{1P}) and stars with correlated enhancements in nitrogen, sodium and aluminum, and depletions in carbon, oxygen and magnesium (the \textit{second population}, or \textit{2P}). 
\citet{Carretta2009} and \citet{Meszaros2015} are both broad studies that spectroscopically map out this phenomenon in the Milky Way.
There are a small number of GCs for which no 2P stars have been identified, including Ruprecht 106 and E3 in the Milky Way (\citealt{Villanova2013}; \citealt{Salinas2015}) and NGC 419 in the Small Magellanic Cloud \citep{Martocchia2017}. 
Although this abundance signature is both ubiquitous and unique in GCs, none of the mechanisms proposed for its origin are able to explain all of the observed properties of GCs (e.g. \citealt{Bas_Lard}, \citealt{Gratton2019}, \citealt{Milone2022}).

One clue to the origin of these systems is the relationship between the mass of the cluster and the fraction of 2P stars found therein.
Previous work has 
found that 
more massive clusters contain a higher fraction of 2P stars than smaller clusters (e.g. \citealt{Bas_Lard}, using 2P star fractions from \citealt{Milone2017} and masses from \citealt{Baumgardt2018}).
This correlation is even more apparent when comparing the 2P star fraction to the calculated \textit{initial mass} of a globular cluster, instead of comparing to the present-day mass \citep{Gratton2019, Usman2024}. 
There appears to be an initial mass threshold, in the range $10^{4.7-5.0}$\msun, that a globular cluster must achieve in order to contain 2P stars.
Once the initial mass passes this critical mass, clusters can achieve much higher enriched star fractions: GCs with initial masses around $10^{5.5}$\msun and $10^{6.2}$\msun contain 2P star fractions of approximately 50\% and 75\% respectively.

We can probe the low-mass end of this relation and further constrain the initial mass threshold by observing and analyzing the chemistry of low-mass, disrupting clusters, or that of fully disrupted clusters in the form of stellar streams.
The \sfive collaboration has done one such analysis on the stellar stream 300S \citep{Usman2024}. 
This globular cluster remnant had an estimated stream mass of about $10^{4.5}$\msun, and one star in eight with an enrichment pattern consistent with 2P stars, significantly lower than the 2P fraction in intact Milky Way GCs of comparable current mass. 
However, the upper limit for the initial mass of the system was calculated to be $10^{4.9}$\msun.
The enrichment of the stream agreed well with systems of comparable initial mass when compared using the stream progenitor's initial mass.

The exact mechanism that links the mass of a cluster and its enriched star fraction is obfuscated because of the complexities in calculating initial mass.
Current calculations of globular cluster mass loss rely on relations inferred from N-body simulations, e.g. those performed in \citet{Baumgardt2003, Lamers2005} and \citet{Baumgardt2018}.
In addition to this, the spatial distribution of 1P and 2P stars could impact the measured star fraction.
For example, \citet{Lardo2011} found the clusters M2, M3, M5, M13, M15, M92 and M53 have more 
2P stars towards their centers, though more recent surveys suggest M5 \citep{Lee2017} and M15 \citep{Nardiello2018} are more mixed than previously thought.
A few clusters have shown evidence of dynamical mixing, with unenriched and enriched populations becoming spatially distributed throughout the cluster (e.g. NGC 6752, by \citealt{Milone2013}; NGC 6362, by \citealt{Dalessandro2014}).
If a cluster is not sufficiently mixed prior to experiencing mass loss, 1P stars may be preferentially stripped from the cluster \citep{Vesperini2021}.
The resulting cluster would have a heightened enriched star fraction relative to its current mass.

While GCs can have large fractions of 2P stars, which are moderately enriched in nitrogen and depleted in carbon, there exists a distinct category of stars which are highly enriched in nitrogen, referred to as \textit{nitrogen-enriched metal-poor} (NEMP) stars.
NEMP stars are defined by \citet{Johnson2007} as having [N/Fe] $> +0.5$ and [C/N] $< -0.5$, while at low metallicity, [Fe/H] $< -2.0$.
Unlike 2P stars, NEMP stars are thought to be enriched via accretion from a companion star on the asymptotic giant branch (AGB, \citealt{Aoki2007, Pols2012}).
Evidence for this enrichment mechanism is supported by the common co-identification of \textit{s}-process elements, such as barium, strontium and yttrium which are synthesized by AGB stars \citep{Luck1991, Izzard2009}.
Stars with similar nitrogen and \textit{s}-process enhancement can sometimes be referred to as barium-enhanced or Ba stars (e.g. \citealt{Luck1991, DOrazi2010}) and carbon-enhanced or CH stars (e.g. \citealt{Cote1997}).

Since NEMP/Ba/CH stars are thought to form in binaries, GCs may not be an ideal location for the evolution of these stars, as the high density of stars in GCs causes binary systems to be disrupted \citep{Cote1997}.
Indeed, \citet{DOrazi2010} observed 1,205 red giant branch stars in GCs and identified just five Ba stars, a rate of 0.4\%, compared to the 2\% rate of Ba stars among field stars \citep{Luck1991}.
The authors also suggest the higher rate of Ba stars among 1P stars relative to 2P stars is indicative of the environment in which 2P stars form: a denser environment would both allow for 2P stars to form and disrupt binary systems, causing the rates of 2P stars and Ba stars to be anti-correlated.

Similarly, a Ba-enriched star has been identified in the aforementioned globular cluster stellar stream 300S \citep{Usman2024}.
The identification of a mass-transfer star in a stellar stream could indicate a disrupting or fully disrupted system could be a more hospitable environment for the long-term survival of binary stars.
By exploring and chemically tagging stars in such systems, we can further probe the relationship between these GCs and their enriched star populations.

ESO~280-SC06 is one of the faintest and most metal-poor GCs in the Milky Way currently known. 
ESO~280-SC06 was discovered by the ESO/Uppsala survey of the southern sky and was originally identified as an open cluster \citep{Holmberg1977, Lauberts1982}.
The cluster was classified as a GC several decades later by \citet{eso_2000}, and was found to have a metallicity of \feh = $-$1.8 and a distance of 21.9 kpc.
The metallicity of the cluster was re-measured as part of a survey of 153 clusters observed by \citet{Bica2006} and was found using photometry from La Silla telescope to have a metallicity of \feh = $-$2.00 and a distance of 21.7 kpc from the sun.
The cluster was studied again using 2MASS photometry by \citet{Bonatto2008} and found to have a metallicity of \feh = $-$1.8 and a total magnitude of $M_V$ = $-$4.9.
The authors suggest this total magnitude should be treated as a lower limit, as the analysis does not take into account different spectral types of stars and ESO~280-SC06 is particularly far at a distance of $\sim$22 kpc.
The magnitude was later remeasured by \citet{Baumgardt2020} using the Hubble Space Telescope and ground-based photometry and was indeed found to be brighter than the previous estimate, at a magnitude $M_V$ = $-$4.28.

\citet{Simpson2018} observed ESO~280-SC06 using the 2dF/AAOmega spectrograph on the Anglo-Australian Telescope and identified 13 red giant branch member stars.
The author used calcium triplet line strengths to infer a cluster metallicity of [Fe/H]=$-2.48^{+0.06 }_{ -0.11}$ and measured a mass of $10^{4.1\pm0.1}$\msun.
These cluster parameters were later revised again by \citet{Simpson2019}, who observed and identified 23 member stars in the cluster.
The authors measured a metallicity of \feh = $-$2.47, found the cluster's distance to be 20.6 kpc from the Sun and identified an NEMP star.
The distance to the cluster was later revisited by \citet{Baumgardt2021}, who found the distance to be 20.95 $\pm$ 0.66 kpc, in agreement with \citet{Simpson2019}.
\citet{Massari2019} combined kinematic data from \textit{Gaia} with estimated cluster ages for Milky Way GCs and analyzed their integrals of motion in order to identify common origins.
The authors suggest that ESO~280-SC06 originally formed in the Gaia-Sausage-Enceladus (\textit{GSE}, \citealt{Belokurov2018, Helmi2018}), and fell into the Milky Way with its host galaxy.
There are currently no estimates of multiple population fractions in ESO~280-SC06.

In this paper, we observe and analyze 10 red giant branch stars in the cluster to estimate its 2P star fraction.
Here we present the chemical abundances of these 10 stars with high-resolution spectroscopy using the Magellan telescopes. 
In Section \ref{sec:obs}, we detail our observations.
Section \ref{sec:chem_abun} describes our chemical abundance analysis methods and results. 
In Section \ref{sec:kinematics}, we report a kinematic analysis of ESO~280-SC06. 
We discuss our results in Section \ref{sec:discussion} and conclude in Section \ref{sec:conclusion}. 

\begin{figure*}[ht]
\centering
\includegraphics[scale = 0.6]{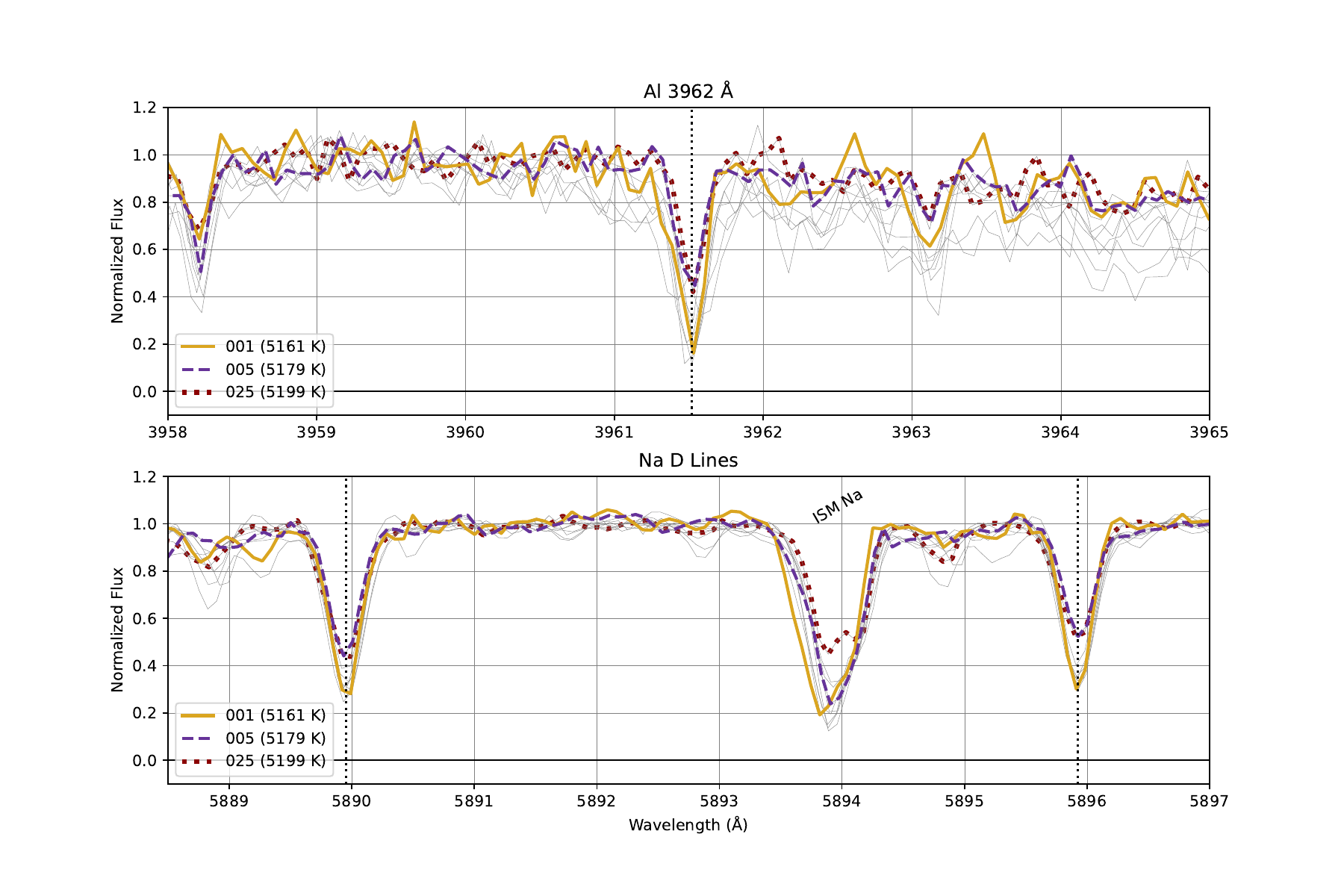}
\vspace{-1.3cm}
\caption{The normalized spectra of the ESO~280-SC06 stars around the measured sodium doublet at 5890 and 5896 \AA~(top) and the measured aluminum line at 3962 \AA~(bottom).
The thin, dotted vertical lines indicate the precise locations of the sodium lines and aluminum line in their respective panels.
Stars 001, 005 and 025 are represented by a gold solid line, a blue dashed line, and a dotted red line, respectively.
All stars have a temperature around 5180 K.
Despite the similar temperatures, stars 005 and 025 clearly display much weaker sodium and aluminum absorption lines.
This demonstrates that both stars are 1P stars, while star 001 is a 2P star.
The other stars' spectra are shown as thin gray lines in the background.
In the bottom panel, \textit{ISM Na} indicates an absorption feature in the spectrum due to sodium in the interstellar medium.}
\label{fig:spec}
\end{figure*}

\begin{figure}[th]
\includegraphics[scale = 0.5]{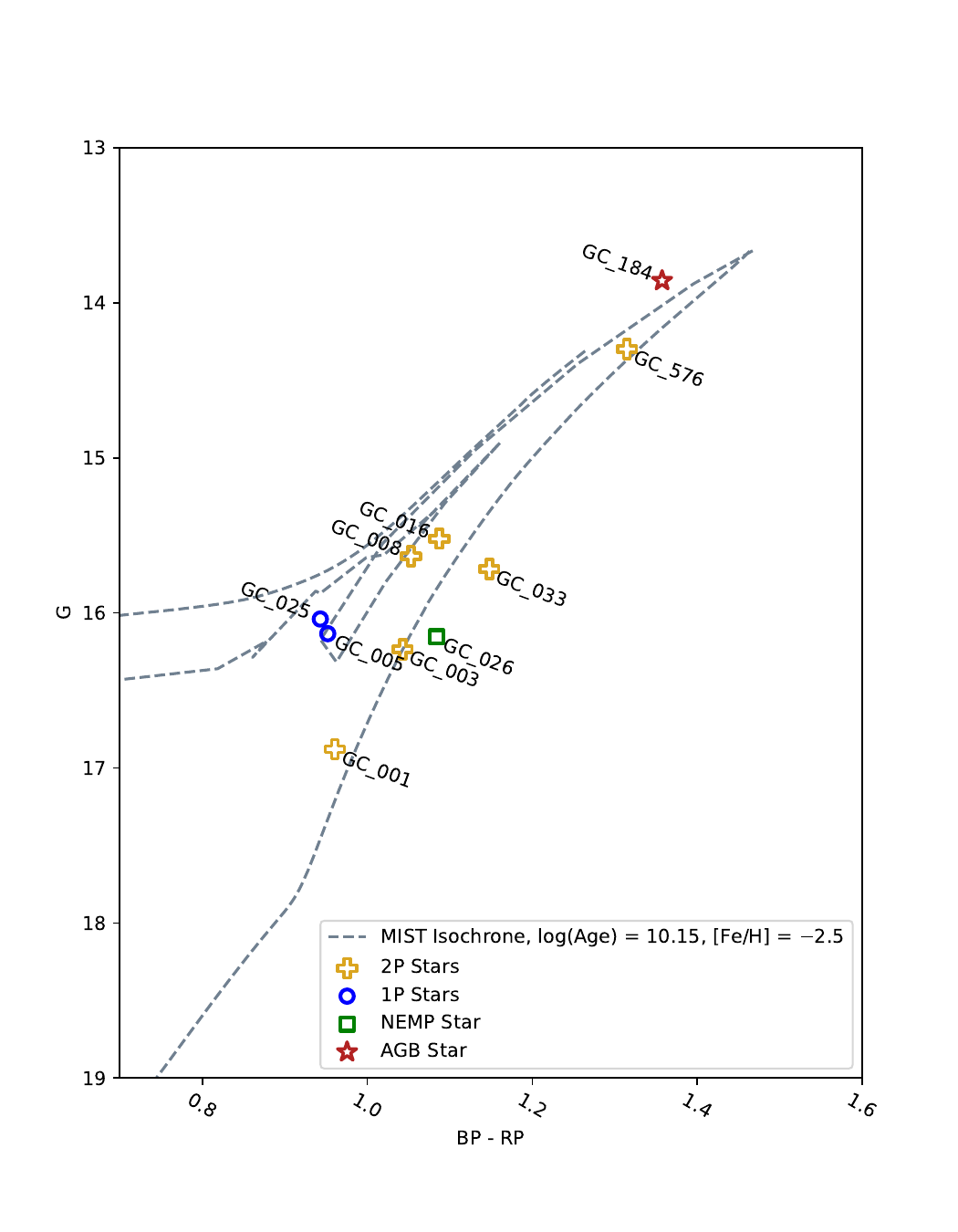}
\caption{The color-magnitude diagram for ESO~280-SC06 stars.
The dotted line represents a MIST isochrone with $\log$ age = 10.15 and metallicity \feh = $-2.5$ \citep{Dotter2016, Choi2016}.
We shift the isochrone to the red side by 0.08 (as was done in previous analyses such as \citealt{Simpson2018}).
The first-population / 1P stars, AGB star, NEMP star and remaining second-population / 2P stars are denoted to blue hollow circles, a red hollow star, a green hollow square and a yellow hollow plus symbols.}
\label{Figure:cmd}
\end{figure}

\section{Observations and Data Reduction} \label{sec:obs}

\renewcommand{\arraystretch}{1.5} 
\begin{table*}[t]
\centering
\caption{Summary of ESO~280-SC06 Properties}
\label{table:cluster_obs}
\begin{tabular}{|cccc|}
\hline\hline
Variables& & Values & Source \\
\hline
RA &(h:m:s) & 18:09:06.0 & \citet{Harris96} \\
Dec &(d:m:s) & $-$46:25:23 & \citet{Harris96} \\ 
Distance &(kpc) & 20.95 $\pm$ 0.66 & \citet{Baumgardt2021} \\
Proper Motion RA &(\masyr) & $-$0.55 & \citet{Simpson2019} \\
Proper Motion Dec &(\masyr) & $-$2.69 & \citet{Simpson2019} \\
Radial Velocity &(\kms)& 94.9 $\pm$ 0.5 & \citet{Simpson2019} \\
M$_{\text{V}}$ & & -4.28 & \citet{Baumgardt2020} \\
Core Radius &(pc) & 1.5 $\pm $0.0 & \citet{Simpson2018} \\
Tidal Radius &(pc) & 53.1 $\pm$ 15.0 & \citet{Simpson2018} \\
$M_{\text{fin}}$ & (\msun) & $10^{4.0 - 4.2}$ & \citet{Simpson2018} \\
\hline
$M_{\text{ini}}$ & (\msun) & $10^{5.4 - 5.7}$ & \\
Apocenter & (kpc) & 13.8 & \\
Pericenter & (kpc) & 1.23 & \\
Eccentricity & & 0.84 & \\
\feh & & $-$2.54 $\pm$ 0.06 & \\
2P Fraction & & $0.80^{+0.07}_{-0.18}$ & \\
\hline
\end{tabular}
\tablecomments{\vspace{0.1cm}The distance, proper motions, radial velocity, current mass $M_{\text{fin}}$ and radius measurements are from literature sources.
We discuss our measured metallicity and 2P star fraction in Section~\ref{sec:chem_abun}, Subsection~\ref{sec:abun_results}.
The orbital parameters are discussed in Section~\ref{sec:kinematics} and we calculate the initial mass in Subsection~\ref{sec:mass}.}
\end{table*}  

We selected ten of the brightest member stars of ESO~280-SC06  from \citet{Simpson2018} and \citet{Simpson2019} to observe with Magellan/MIKE \citep{Bernstein2003}.
Observations were conducted on March 9, June 23-24, and July 25-27 in 2017.
Data from adjacent nights were reduced together with CarPy \citep{Kelson2003}.
All 10 stars were observed with a 0.7$^{\prime \prime}\times$5.0$^{\prime \prime}$ slit, with 2$\times$2 on-chip binning to reduce read noise.
The slit yields a spectral resolution of $\sim$28,000 and $\sim$35,000 in the red and blue, respectively. 
We can see an example of these spectra in Figure~\ref{fig:spec}, which shows the measured sodium doublet and aluminum line.
Each order of each reduced spectrum was normalized using a third degree natural spline. 
Radial velocities are measured by cross-correlating the Mg b region with a high signal-to-noise spectrum of HD122563.
After correcting for the radial velocities, data from different runs are combined order-by-order using an inverse-variance weighted average of the individually normalized spectra, then stitched into a single spectrum\footnote{Code available at \url{https://github.com/alexji/alexmods/blob/master/alexmods/specutils/continuum.py}}.
Table~\ref{table:cluster_obs} details the cluster parameters and Table~\ref{table:stars_obs} details each star's observations.  

\begin{table*}[ht!]
\caption{Summary of Observed Stars}
\label{table:stars_obs}
\begin{tabular}{|ccccccccccc|} 
\hline\hline
Star & Gaia ID &  \gaia G & $t_{exp}$ & SNR & SNR & \vhel & \teff & \logg & \vt  & \feh \\ 
     &         & (mag)    &  (min)    & 4500\AA~  &  6500\AA~ & (\kms) & (K)   & (dex) & (\kms) & (dex)   \\ 
\hline
001~ & 6719598998858700032  & 17.22 & 120 & 16 & 34 & 95.0 & $5161\pm~~75$  & $2.19 \pm 0.15$ & $1.48 \pm 0.2$ & $-2.50 \pm 0.2$ \\ 
003~ & 6719599101937916544  & 16.58 & 171 & 16 & 43 & 93.0 & $4988\pm~~75$  & $1.86 \pm 0.15$ & $2.27 \pm 0.2$ & $-2.68 \pm 0.2$ \\ 
005$^\dagger$ & 6719599209329902720  & 16.48 & 100 & 20 & 44 & 93.1 & $5179\pm~~75$  & $1.90 \pm 0.15$ & $1.82 \pm 0.2$ & $-2.48 \pm 0.2$ \\ 
008~ & 6719598998858703744  & 15.98 & 70  & 18 & 42 & 90.9 & $4967\pm~~75$  & $1.60 \pm 0.15$ & $2.08 \pm 0.2$ & $-2.47 \pm 0.2$ \\ 
016~ & 6719599174970284928  & 15.86 & 60  & 21 & 47 & 93.4 & $4900\pm~~75$  & $1.54 \pm 0.15$ & $1.89 \pm 0.2$ & $-2.50 \pm 0.2$ \\ 
025$^\dagger$ & 6719599170657398400  & 16.39 & 134 & 19 & 44 & 92.6 & $5199\pm~~75$  & $1.87 \pm 0.15$ & $2.39 \pm 0.2$ & $-2.54 \pm 0.2$ \\ 
026$^{\ddagger}$ & 6719598900092253184  & 16.51 & 139 & 20 & 50 & 91.9 & $4907\pm~~75$  & $1.79 \pm 0.15$ & $1.83 \pm 0.2$ & $-2.38 \pm 0.2$ \\ 
033~ & 6719598174224943104  & 16.07 & 75  & 19 & 46 & 91.9 & $4783\pm~~75$  & $1.57 \pm 0.15$ & $2.16 \pm 0.2$ & $-2.59 \pm 0.2$ \\ 
184$^*$ & 6719599003157597184  & 14.16 & 15  & 20 & 58 & 89.0 & $4441\pm~200$ & $0.65 \pm 0.15$ & $2.13 \pm 0.2$ & $-2.58 \pm 0.2$ \\ 
576~ & 6719598075458648576  & 14.64 & 40  & 28 & 70 & 93.0 & $4506\pm~~75$  & $0.86 \pm 0.15$ & $2.83 \pm 0.2$ & $-2.69 \pm 0.2$ \\ 
\hline
\end{tabular}
\tablecomments{The stellar parameters of observed stars in this analysis.
All stars are red giant branch stars, with the probable exception of star 184 (denoted by $^*$), which appears to be an asymptotic branch star.
The effective temperatures $T_{\text{eff}}$ are calculated using the relations described in \citet{Mucciarelli2021}.
Star 184 returns three extremely different temperatures from these relations, depending on which color is used for the calculation.
$\logg$ is calculated using the relation in \citet{Ji2020b}. The microturbulence velocity and metallicity are calculated in our spectroscopic analysis.
Stars 005 and 025 (denoted with $^{\dagger}$) are the identified 1P stars.
Star 026 (denoted with $^{\ddagger}$) is the NEMP star identified in \citet{Simpson2019}.}
\end{table*}   

\begin{table*}[ht]
\caption{Example Line Table, Star 576}
\label{table:line_table}
\begin{tabular}{|ccccccccccccccc|} 
\hline \hline
Element & Wavelength (\AA) & expot & loggf & & log $\epsilon$ & $e_{\text{stat}}$ & eqw & $e_{\text{eqw}}$ & FWHM & $e_{\text{Teff}}$ & $e_{\log g}$ & $e_{\nu t}$ & $e_{\text{MH}}$ & $e_{\text{tot}}$ \\
\hline 
... & ...  & ...     & ...   & ...     & ~   & ...     & ...   & ...     & ...   & ...   & ...     & ...     & ...  & ...    \\
O I & 6300.0 & 0.0 & $-$9.82 & $<$ & 7.069 & $-$ & $-$ & $-$ & $-$ & $-$ & $-$ & $-$ & $-$ & $-$ \\
Fe I & 6400.001 & 3.603 & $-$0.27 & & 5.073 & 0.071 & 43.463 & 4.569 & 0.385 & 0.058 & $-$0.006 & $-$0.015 & 0.003 & 0.14 \\
Fe I & 6430.846 & 2.174 & $-$1.95 & & 5.124 & 0.082 & 39.666 & 5.224 & 0.325 & 0.081 & $-$0.004 & $-$0.013 & 0.003 & 0.15 \\
Ca I & 6439.075 & 2.524 & 0.47 & & 4.248 & 0.081 & 72.584 & 5.399 & 0.443 & 0.041 & $-$0.013 & $-$0.037 & 0.002 & 0.14 \\
K I & 7664.911 & 0.0 & 0.125 & & 3.279 & 0.047 & 66.39 & 3.393 & 0.365 & 0.049 & $-$0.012 & $-$0.032 & 0.002 & 0.123 \\
K I & 7698.974 & 0.0 & $-$0.178 & & 3.292 & 0.064 & 46.477 & 4.298 & 0.444 & 0.052 & $-$0.008 & $-$0.018 & $-$0.001 & 0.13 \\
Sc II & 5030.0 & $-$ & $-$ & & 0.684 & 0.294 & $-$ & $-$ & 0.188 & 0.04 & 0.054 & $-$0.005 & 0.006 & 0.32 \\
Sc II & 4669.0 & $-$ & $-$ & & 0.843 & 0.139 & $-$ & $-$ & 0.148 & 0.031 & 0.053 & $-$0.006 & 0.006 & 0.18 \\
... & ...  & ...     & ...   & ...     & ~   & ...     & ...   & ...     & ...   & ...   & ...     & ...     & ... & ... \\
\hline
\end{tabular}
\tablecomments{This table is available in its entirety in machine-readable form as part of the arXiv source.
expot is excitation potential; loggf is oscillator strength; log $\epsilon$ is the absolute stellar abundance to H; $e_{stat}$ is the statistical abundance uncertainty due to the equivalent width uncertainty; eqw is equivalent width; $e_{eqw}$ is uncertainty in equivalent width; FWHM is the full width of the line at half maximum; $e_{T_{\text{eff}}}$ is the difference on the abundance due to $1\sigma$ uncertainties on effective temperature, and similarly for $e_g$ for the surface gravity, $e_{v}$ for the microturbulence, and $e_{M}$ for the metallicity.
Elements with no listed equivalent width were measured using synthesis.
The wavelengths of synthesized lines have been rounded to the nearest whole wavelength.
Elements marked with the less-than symbol, $<$, are 3$\sigma$ upper limits.}
\end{table*}

\section{Chemical Abundance Analysis} \label{sec:chem_abun}

\subsection{Stellar Parameters}


For our analysis, we estimate the stellar parameters using \textit{Gaia} photometry.
First, we dereddened using Equation~1 from \citet{Babusiaux2018}. 
We used reddening data from \citet{Schlegel1998} with corrections from \citet{Schlafly11}. 
We accessed the E(B--V) from the online database IRSA.\footnote{\url{https://irsa.ipac.caltech.edu/applications/DUST/}}
We then used the dereddened photometric data to estimate the \teff and \logg 
using Equation~(1) from \citet{Mucciarelli2021} for BP$-$RP and Equation~(3) from \citet{Ji2020b}, respectively.

We note that the photometric temperature for star 184 was more sensitive than the other stars to the choice of color used in the photometric relation from \citet{Mucciarelli2021}.
We suspect this star is an AGB, and therefore does not exactly follow predicted temperatures for RGB stars described in the relations.
We subsequently chose to use the temperature resulting from BP$-$RP, as this was both the median photometric temperature and was consistent with the color relation used for the other observed stars.
To account for our uncertainty on temperatures in AGB stars relative to predictions for RGB stars, we use an increased uncertainty of 200 K for star 184, as is reflected in Table~\ref{table:stars_obs}.

\subsection{Abundance Calculations}

The overall analysis steps follow the method performed in \citet{Atzberger2024}.
The majority of the analysis was conducted using the code \code{SMHR} (first described in \citealt{Casey_2014}, most recently in \citealt{Casey2025}), 
which provides a graphical user interface to fit equivalent widths, synthesize more complex absorption regions, interpolate ATLAS model atmospheres \citep{Castelli2004}, run MOOG to determine chemical abundances through curves of growth \citep{Sneden1973, Sobeck2011}, and calculate stellar parameter and chemical abundance uncertainties \citep{Ji2020b, Atzberger2024}.
The uncertainty for each individual measurement is calculated by propagating uncertainties in the stellar parameters and in the equivalent width measurements through the abundance inference.
An example of a line list and measured uncertainties are presented in Table~\ref{table:line_table}
The overall abundance uncertainty for a star combines these individual uncertainties through a weighted average.
For more details, see \citet{Atzberger2024}.
The analysis is performed assuming local thermodynamic equilibrium (\textit{LTE}), though for two elements we calculate additional corrections to account for non-LTE effects at the end of the analysis.

\subsection{Measurements} \label{sec:abun_results}

Chemical abundances are measured using a combination of equivalent widths and synthesis.
Our results are detailed in Tables \ref{tab:001}--\ref{tab:576}.
We include nLTE abundances for Na and Al in these tables.

\emph{Metallicity.}
Metallicity was determined using Fe I lines, which are measured using equivalent widths.
We use the nested sampling algorithm \code{dynesty} to calculate Bayesian posteriors for the cluster's mean metallicity and metallicity dispersion \citep{Speagle2020, Higson2019, Koposov2022}.
This method is identical to the method used for the abundance dispersion analysis in \citet{Usman2024} (Section~4.1), including the likelihood function (Equation~4).
We find a mean metallicity for the entire cluster of [Fe I/H] = $-$2.54 $\pm$ 0.06, with an upper limit dispersion of 0.09.
Similarly, the mean metallicity for the entire cluster with Fe II is [Fe II/H] = $-$2.55 $\pm$ 0.05, with an upper limit dispersion of 0.08.
This metallicity is consistent with previous work, such as \citet{Simpson2019}.

\emph{Carbon and Nitrogen.}\label{C_enriched} 
Carbon and nitrogen are measured 
through synthesis of CH and CN molecular bands, respectively. 
For the CH bands, we measure the regions around 4310~\AA~and 4323~\AA.
For the CN bands, we measure the region around 3877~\AA.
These bands have significant absorption that are too complex to be modeled with an equivalent width, so we therefore synthesize these regions.
For a few stars, there is no significant observable enrichment, and we instead estimate upper limits.


\emph{$\alpha$ elements.}\label{alpha_elem}
The $\alpha$ elements measured here are magnesium, silicon, calcium and titanium.
Magnesium was measured using equivalent widths of five lines between 4500 and~5600~\AA.
Silicon was measured using an equivalent width at the wavelength 4102~\AA~and with a synthesis at 3906~\AA.
Calcium was measured using equivalent widths of a dozen absorption lines between 4200 and 6500~\AA.
Titanium I was measured using equivalent widths of six lines between 4500 and 5100~\AA.
Titanium II was measured using equivalent widths of about 20 lines between 4000 and 4800~\AA.

\begin{figure*}[t!]
\centering
\includegraphics[scale=0.45]{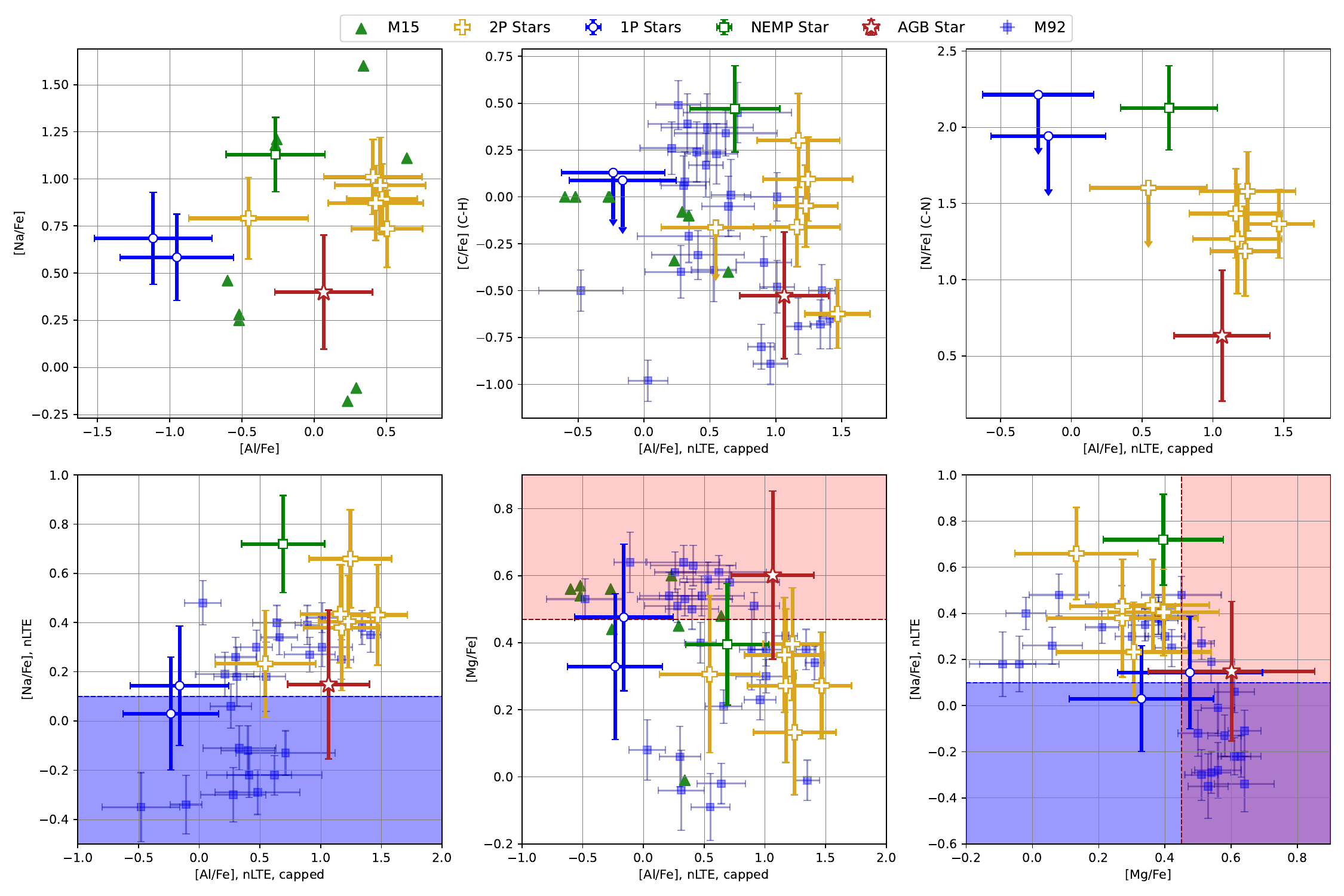}
\caption{Key abundances for identifying chemical patterns found in multiple populations.
2P stars are consistently enriched in sodium, aluminum, and nitrogen, and depleted in magnesium and oxygen.
The top left plot compares the LTE sodium and aluminum abundances.
Directly below this, the bottom left plot adjusts these abundances to incorporate nLTE corrections.
The top center, top right and bottom center plots compare the carbon abundances (measured using the molecular C-H band), nitrogen abundances (measured using the molecular C-N band) and magnesium abundances, to the nLTE aluminum abundances.
Lastly, the bottom right plot compares the nLTE sodium abundances to magnesium abundances.
We identify eight stars with chemical enrichment patterns indicative of second-population/2P stars.
Stars 005 and and 025 (represented by hollow blue circles) are the only stars that have low sodium and aluminum abundances, and are therefore classified as first-population/1P stars.
Star 184, which has been identified as an AGB, is represented by a red hollow star.
The NEMP star (identified in \citealt{Simpson2019}) is represented by a green hollow square.
The remaining second-population stars are represented by dark yellow plus symbols.
For comparison, we also show abundances from M15 \citep{Sobeck2011}, represented by solid green triangles, and M92 \citep{Kirby2023}, represented by solid blue squares with error bars.
For nLTE-corrected sodium abundances and magnesium abundances, we overlay a blue region and a light red hatched region to represent the diagnostic thresholds used in \citet{Kirby2023} to classify 1P stars: [Na/Fe] $<$ 0.1 and [Mg/Fe] $>$ 0.45 indicate a star is 1P.
The authors suggest the sodium threshold is more reliable than the magnesium threshold.
By these categorizations, 005 is the only star that is sodium-poor enough to definitively be considered 1P.
Star 025, however, has very similar sodium and aluminum abundances, and is likely also a 1P star, despite being slightly over the threshold.
Star 184 is magnesium-rich enough to be classified as a 1P star, however, its high aluminum abundance is incongruent with being a 1P star; we therefore consider it a 2P star.
\label{fig:multiplepop}}
\end{figure*}

\emph{odd-Z elements.} \label{odd-z_elem}
The odd-Z elements measured here are sodium, potassium and aluminum.
Sodium was measured by synthesizing the sodium D lines at 5890 and 5896~\AA.
Aluminum was measured by synthesizing the absorption line at 3962~\AA.
We also synthesize a large hydrogen line located in the spectrum next to 3962~\AA~to ensure the aluminum line is accurately measured.
While these abundances are measured assuming LTE, we also estimate nLTE corrections for sodium and aluminum, as discussed in Section~\ref{subsec:nlte}.
Potassium is measured using equivalent widths of the lines at 7664 and 7698~\AA.

\emph{Iron-peak elements.}\label{fe-peak_elem}
Iron-peak elements measured here are chromium, manganese, nickel and zinc.
Chromium was measured using equivalent widths at the lines 4652 and 5409~\AA.
Manganese was measured by synthesizing three lines at wavelengths 4824, 4782, and 4753~\AA.
Note that we exclude the manganese triplet around 4031$-$4033~\AA, as we expect that there are unaccounted uncertainties, perhaps due to nLTE effects.
Nickel is measured using two equivalent widths at 4401 and 4714~\AA, while the line at 5476~\AA~is synthesized.
Lastly, zinc is measured using a synthesized line at 4811~\AA.

\emph{Neutron-capture elements.}\label{neutron_cap}
The neutron-capture elements measured are strontium, yttrium, zirconium, barium, lanthanum, europium and dysprosium.
Strontium II is measured by synthesizing 4078 and 4215~\AA.
Yttrium is measured by synthesizing lines at 4900, 4884 and 4397~\AA.
Zirconium is measured by synthesizing the line at 4208~\AA.
Barium is measured by synthesizing five lines between 4500 and 6500~\AA.
Lanthanum is measured by synthesizing four lines between 4000 and 5000~\AA. 
Europium is generally too low to be measured 
using the line at 6645~\AA, so in some stars, we synthesize up to three bluer lines between 4100 and 4450~\AA.

\begin{table} [h]
\caption{nLTE Corrections for aluminum and sodium.}
\centering
\begin{tabular}{c|c|c} 
\hline\hline
Star    & Na & Al \\ 
\hline
001 & $-$0.3510~ & +0.839~~\\
003 & $-$0.5865~ & +0.717~~\\
005 & $-$0.5540~ & +0.717~~\\
008 & $-$0.5585~ & +1.131$^{\dagger}$~\\
016 & $-$0.4880~ & +0.759~~\\
025 & $-$0.5410~ & +0.953~~\\
026 & $-$0.4095~ & +0.958~~\\
033 & $-$0.4365~ & +0.737~~\\
184 & $-$0.2495* & ~+1.187*$^{\dagger}$\\ 
576 & $-$0.3040* & +0.965*~\\ 
\hline
\end{tabular}
\tablecomments{For the two coldest stars (marked with an asterisk *), the calculated surface gravity was less than one, $\logg<1$.
The nLTE calculators leverage sets of grids of stellar atmospheres which span a wide range of stellar parameters.
The grids do not extensively cover surface gravities below $\logg<1$.
We therefore round the surface gravity of the stars 184 and 576 up to \logg=1.
The calculated Al correction for stars 008 and 184 (denoted with a dagger $\dagger$) are higher than 1; however, \citet{Nordlander_2017} suggests Al corrections above 1 are not reliable, and should instead be capped at 1.
Despite using these reduced corrections, both stars have very high Al abundances and are therefore still classified as 2P stars.
\label{table:nLTE}}
\end{table}  

\subsection{nLTE Corrections}
\label{subsec:nlte}

%

In the bulk of our analysis, we assume our stars are in local thermodynamic equilibrium, \textit{LTE}.
However, this assumption is often violated in 
massive stars.
Some inferred abundances can therefore be over- or underestimated, and therefore we apply non-LTE (\textit{nLTE)} corrections to account for this discrepancy.

The abundances of two key elements characteristic of multiple populations, aluminum and sodium, can often be misestimated due to nLTE effects.
We determined nLTE abundance corrections for our stars from \citet{Lind_2011} for sodium and \citet{Nordlander_2017} for aluminum.
Table~\ref{table:nLTE} provides aluminum and sodium corrections for each star.
For sodium we used the average correction between lines 5890 and 5896~\AA~with a range of $-0.6$ to $-0.2$.

We used the line at 3961~\AA~for aluminum, which had a correction range of $+0.7$ to $+1.2$.
However, \citet{Nordlander_2017} suggests that nLTE corrections may reach only as large as +1 dex.
Corrections above that range may not be reliable above +1 dex, due to limitations in the grid models \citep{Nordlander2019}.
For this reason, we cap nLTE aluminum abundances to +1 dex if that limit was exceeded, as was the case for stars 008 and 184.
Despite these capped corrections, both stars are very aluminum rich and are still classified as 2P stars.

We note that for some stars, the stellar parameters needed to be adjusted slightly to fit within the target parameter range to estimate the nLTE corrections.
Stars 184 and 576 were estimated using a \logg of 1, instead of their true value.
These adjustments may lead to a small error in these stars' corrections.

\renewcommand{\arraystretch}{1.5} 
\begin{table}
    \centering
    \caption{Cluster Abundance Means \& Dispersions}
    \begin{tabular}{ccc}
Element  &  Mean Abundance  &  Dispersion \\
\hline
[Fe/H]~~ & $-2.54\pm~0.06$ & $<$0.11 \\\relax
[Fe II/H]~~ & $-2.55\pm~0.05$ & $<$0.10 \\
\hline
[C/Fe I]$^{\dagger}$~ & $-$0.17 $\pm 0.14$ & 0.20$^{+0.20}_{-0.15}$ \\\relax
[N/Fe I]$^{\dagger}$~ & ~~0.95 $^{+0.04}_{-0.11}$ & 0.29$^{+0.27}_{-0.23}$ \\\relax
[Na/Fe I]~~ & ~~0.84 $\pm~0.07$ & $<$0.18 \\\relax
[Na/Fe I]$_{\text{nLTE}}$ & ~~0.39 $\pm~0.07$ & $<$0.21 \\\relax
[Mg/Fe I]~~ & ~~0.34 $\pm~0.06$ & $<$0.13 \\\relax
[Al/Fe I]~~ & ~~0.03 $\pm 0.19$ & 0.44$^{+0.22}_{-0.19}$ \\\relax
[Al/Fe I]$_{\text{nLTE}}$ &~~ 0.84 $^{+0.11}_{-0.17}$ & 0.41$^{+0.21}_{-0.19}$ \\\relax
[Si/Fe I]~~ & ~~0.59 $\pm~0.09$ & $<$0.14 \\\relax
[K/Fe I]~~ & ~~0.70 $\pm~0.07$ & $<$0.12 \\\relax
[Ca/Fe I]~~ & ~~0.38 $\pm~0.06$ & $<$0.11 \\\relax
[Sc/Fe II]~ & ~~0.10 $\pm~0.06$ & $<$0.12 \\\relax
[Ti/Fe I]~~ & ~~0.28 $\pm~0.07$ & $<$0.13 \\\relax
[Ti II/Fe II]~ & ~~0.38 $\pm~0.07$ & $<$0.12 \\\relax
[V/Fe I]~~ & ~~0.11 $\pm~0.10$& $<$0.19 \\\relax
[Cr/Fe I]~~ & $-$0.17 $\pm~0.07$ & $<$0.15 \\\relax
[Mn/Fe I]~~ & $-$0.38 $\pm~0.08$ & $<$0.14 \\\relax
[Co/Fe I]~~ & ~~0.16 $\pm~0.08$ & $<$0.14 \\\relax
[Ni/Fe I]~~ & ~~0.07 $\pm~0.07$ & $<$0.14 \\\relax
[Zn/Fe I]~~ & ~~0.20 $\pm~0.10$ & $<$0.17 \\\relax
[Sr/Fe II]$^{\dagger}$ & ~~0.21 $\pm~0.10$& $<$0.25 \\\relax
[Y/Fe II]$^{\dagger}$ & $-$0.12 $\pm~0.08$ & $<$0.14 \\\relax
[Zr/Fe II]~ & ~~0.28 $\pm~0.09$ & $<$0.17 \\\relax
[Ba/Fe II]$^{\dagger}$ & $-$0.34 $\pm~0.06$ & $<$0.12 \\\relax
[La/Fe II]$^{\dagger}$ & $-$0.00 $\pm~0.12$ & $<$0.31 \\\relax
[Eu/Fe II]~ & ~~0.38 $\pm~0.08$ & $<$0.18
    \end{tabular}
    \label{tab:cluster_abunds}
    \tablecomments{Estimated abundance means and dispersions for ESO~280-SC06.
    The error bars represent 1$\sigma$ confidence limits.
    Stars which have upper limits for a given abundance are excluded from its mean and dispersion calculation.
    Star 184 is excluded for elements which can be enriched in NEMP stars (denoted with $^{\dagger}$).
    If a dispersion cannot be measured, we give an upper limit at 95\% confidence.
    If the two 1P stars, 005 and 025 are excluded from the analysis, we measure no dispersion in aluminum.
    We find an upper limit of 0.38.}
\end{table}

\begin{figure*}
\centering
\includegraphics[scale = 0.47]{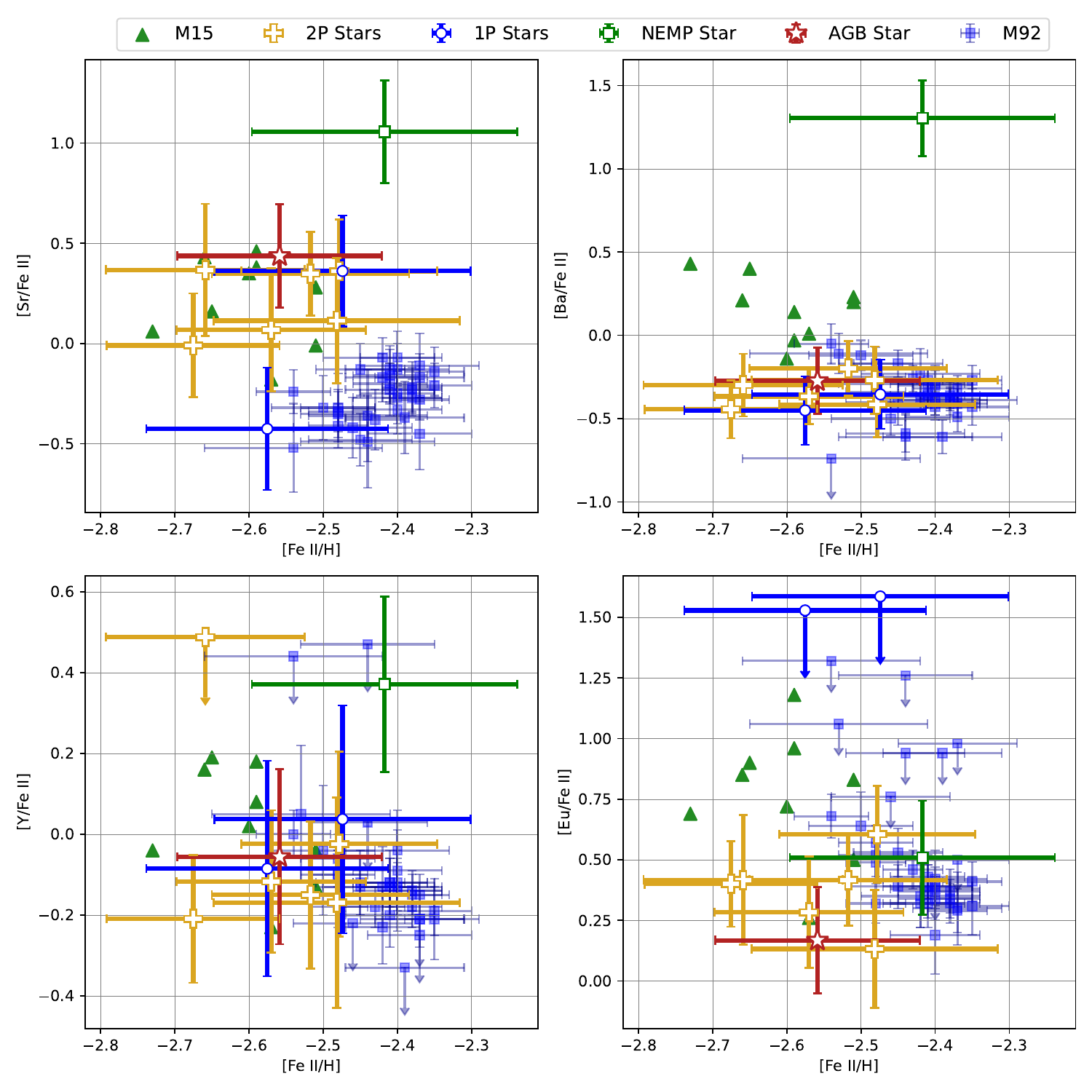}
\caption{Neutron-capture abundances measured in the ESO~280-SC06 stars.
The first-populations/1P stars, the AGB star, the NEMP star, and the remaining second-population/2P stars are represented by blue hollow circles, a red hollow star, a green hollow square and yellow hollow plus symbols, respectively.
For comparison, we also compare to abundances from M15 \citep{Sobeck2011}, represented by green triangles, and M92 \citep{Kirby2023}, represented by blue squares with error bars.
The NEMP star, 026, has significantly higher \textit{s}-process abundances (strontium, barium, and yttrium) than the other stars.
This is consistent with \textit{s}-process enhancement predicted from accretion from an asymptotic giant branch binary companion.
Among the other stars, we find no significant spread in neutron-capture abundances, contrary to patterns previously identified in other systems such as M92 \citep{Kirby2023}.}
\label{fig:ncap}
\end{figure*}

\subsection{Abundance Results}

\subsubsection{Multiple Populations}
In our analysis, we classify eight of the ten observed stars as 2P stars.
The 1P stars are 005 and 025, which show relatively less sodium and aluminum relative to the remaining stars.
In Figure~\ref{fig:spec}, we compare the spectra of 1P stars 005 and 025 (represented by purple dashed and red dotted lines, respectively) to the spectrum of 2P star 001 (represented by a dark yellow solid line) 
in the region 
of the sodium D lines and the aluminum line at 3962 \AA.
All three stars have similar effective temperatures, \teff $\approx 5180$ K.
Despite the similar stellar parameters, the stars 005 and 025 clearly exhibit significantly less absorption at the wavelengths of the aluminum and sodium lines relative to 001, indicating that the former are 1P stars while the latter is a 2P star.

We also compare our measured abundances to those of M92 \citep{Kirby2023} in Figure~\ref{fig:multiplepop}.
In \citet{Kirby2023}, the authors define two thresholds for classifying 1P or 2P stars: stars with nLTE-corrected [Na/Fe] $>0.1$ and/or [Mg/Fe] $<0.45$ are considered 2P stars, though they specify that the constraint on sodium seems to be a more robust classifier.
In our system, stars 005 and 025 show little enrichment in sodium or aluminum, with nLTE-corrected abundances [Na/Fe] = 0.03 and 0.14 and [Al/Fe] = $-$0.23 and $-$0.16, respectively.
Star 025 additionally has high magnesium, further conforming to expected chemical patterns.

A key feature of the multiple population phenomenon is the relative ratios of 1P to 2P stars in clusters, with more massive clusters having higher fractions of 2P stars. 
Specifically, we define the fraction of enriched stars as $f_{\text{enrich}} = \frac{2\rm{P}}{1\rm{P} + 2\rm{P}}$.
We therefore estimate ESO~280-SC06 has an 80\% enrichment fraction, as eight of our ten stars are 2P stars.
We can estimate error bars by assuming the cluster is comprised of two populations with some inherent fraction of enriched stars.
Since there exists no known correlation between a star's magnitude and whether it is a 1P or 2P star \cite[see the color magnitude diagrams from][]{Piotto2015}, we assume our sample of stars is a random draw from a binomial distribution with this endemic enrichment fraction.
We therefore can infer the cluster's enrichment fraction to be $f_{\text{enrich}} = 0.80^{+0.07}_{-0.18}$ using Bayesian statistics.
We discuss further implications of this enrichment fraction with respect to the cluster's mass in Section~\ref{sec:discussion}, Subsection~\ref{subsec:multipop}.

\subsubsection{NEMP Star}
Our analysis also confirms the nitrogen and \textit{s}-process enrichment in star 026, as initially identified by \citet{Simpson2019}.
Figure~\ref{fig:ncap} demonstrates star 026's enrichment in strontium, barium and yttrium relative to the other observed stars.
We also demonstrate that star 026 does \textit{not} present any enrichment in europium, indicating that the star is enriched specifically in \textit{s}-process elements and not \textit{r}-process elements.
This enrichment pattern is consistent with mass transfer from an AGB companion, similar to stars identified in the 300S stream by \citet{Usman2024} and most recently in M55 by \citet{DaCosta2025}.

Although some of the surface abundances of star 026 have been altered by mass transfer, we are confident in our identification of it as a 2P star based on the Na, Mg and Al abundances. The lower-mass AGB stars that are the site of s-process nucleosynthesis are not a major site of the hot hydrogen burning that produces the O-Na and Mg-Al anticorrelations.
We further discuss the implications of finding post-mass transfer stars in GC environments in Section~\ref{sec:discussion}, Subsection~\ref{subsec:CN_star}.

\subsubsection{Neutron-Capture Spread}
If we exclude the NEMP-\textit{s} star, 026, from our analysis, we identify no significant neutron-capture spread in ESO~280-SC06.
This contradicts previous expectations, as similarly metal-poor GCs M15 and M92 have been found to exhibit spreads in these elements \citep{Sobeck2011, Kirby2023}.
Since our abundance uncertainties conservatively include all propagated uncertainties from stellar parameters, our uncertainties may be too large to constrain these element dispersions.
For example, \citet{Kirby2023} constrains a barium dispersion of $0.09^{+0.06}_{-0.05}$.
Among our stars with our conservative uncertainty estimates, we are only able to constrain a dispersion of less than $<0.12$ at 95\% confidence. We therefore cannot achieve the same level of precision presented in \citet{Kirby2023}.

We further discuss the implications of these neutron-capture element spreads in GCs in Section~\ref{sec:discussion}, Subsection~\ref{subsec:ncap}.

\section{Kinematic Analysis} \label{sec:kinematics}

\begin{figure*}
\centering
\includegraphics[scale = 0.55]{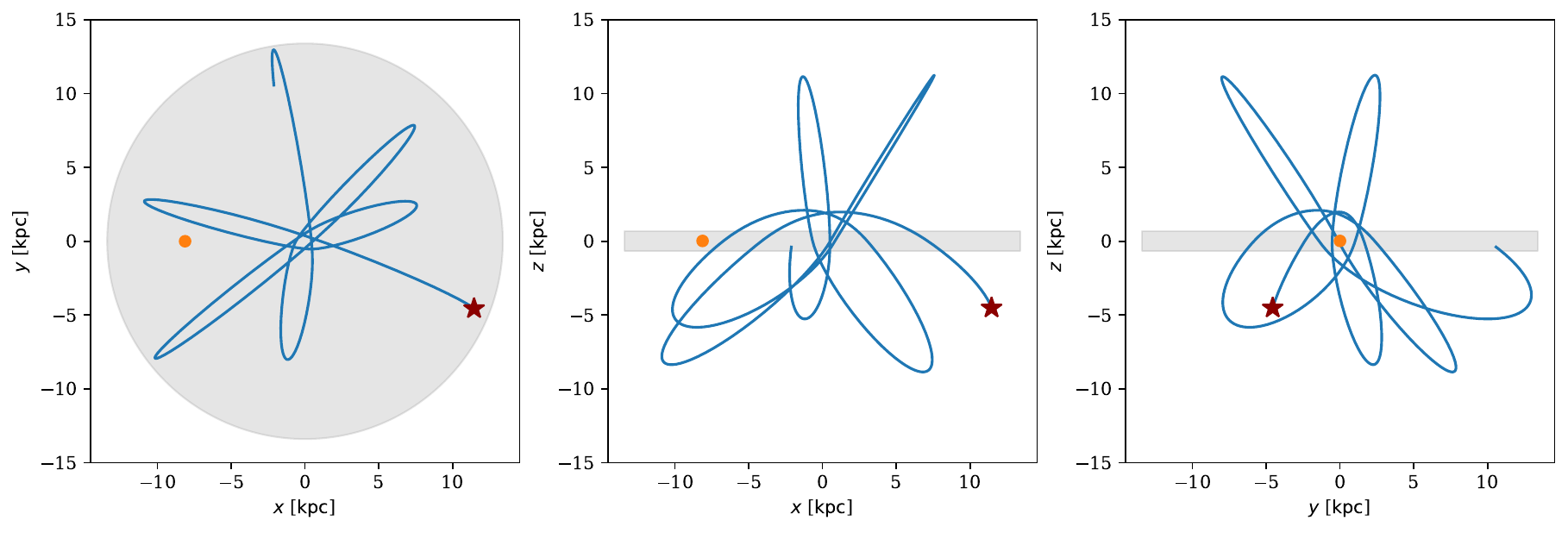}
\caption{Three views of the Galactic orbit of ESO~280-SC06. 
The present position of the cluster is marked with a red pentagon, and its orbit over the past 1 Gyr is shown as a blue line. 
The high eccentricity and inclination of ESO~280-SC06's orbit mean that it has significant disk crossings at each pericenter passage, roughly every 150 Myr.
The position of the Sun is shown with a yellow circle.
The Milky Way is represented by the gray circle in the XY plane, and the thick gray line for side-on views in XZ and YZ planes.
This orbit was integrated using \code{gala} with the default \code{MilkyWayPotential}.}
\label{Figure:orbit}
\end{figure*}

We used \code{gala} \citep{gala,adrian_price_whelan_2021_5057630} to integrate the Galactic orbit and calculate kinematic properties of ESO~280-SC06, using the built-in \code{MilkyWayPotential}. 
The coordinates, proper motion components, and distance are adopted from \citet{Simpson2019} and summarized in Table~\ref{table:cluster_obs}. 
We used a time step of 0.1 Myr and integrated backwards for 1 Gyr. 

The orbit is shown in Figure \ref{Figure:orbit}.
ESO~280-SC06 is currently located on the far side of the Milky Way disk from the Sun.
Its orbit has an apocenter of 13.7 kpc, pericenter of 1.2 kpc, and eccentricity of 0.83. 
The ESO~280-SC06 orbit intercepts the Milky Way disk many times. 
This tight orbit could cause the cluster to lose a significant amount of mass over its lifetime.

\subsection{Initial Mass Estimate}\label{sec:mass}
Based on the orbital parameters, we estimate the mass loss of ESO~280-SC06 in order to estimate the cluster's mass at formation.
We estimate this initial mass using two separate methods:

\vspace{0.4cm}
\emph{Method 1:}
The first method accounts for mass loss due to stellar evolution and tidal dissolution.
We model the globular cluster using a Kroupa initial mass function (\textit{IMF}, \citealt{Kroupa2001}).
We use the parameters for a system with metallicity fraction Z = 0.004 from Table~1 of \citet{Kruijssen2008}.
A metallicity of \feh = $-$2.5 is approximately equivalent to a metallicity of (Z = 0.003), so this approximation is accurate for our purposes.
Given this IMF, the fractional mass loss from stellar evolution $q$, given in Equation~2 in \citet{Lamers2005} as a function of time $t$ is:
\begin{equation}
    \log q(t) = (\log t - 6.9)^{0.256} - 1.696.
\end{equation}
We then use this to estimate the amount of mass still bound in the cluster by Equation~3 of the same paper:
\begin{equation}
    \mu(t) = 1 - q(t)
\end{equation}

We additionally account for the dissolution due to two-body relaxation, as described with Equation 7 of \citet{Kruijssen2009}:
\begin{equation}
    t_0 = t_{0,\odot} \left(\frac{r_{apo}}{8.5 \text{kpc}}\right) \left( \frac{V_c}{220 \frac{km}{s}} \right)^{-1}(1-\epsilon)
\end{equation}
where $t_0$ and $t_{0,\odot}$ are the dissolution times of our system and a system at the Galactocentric distance of the Sun,
$r_{apo}$ is the distance at the orbit's apocenter, $V_c$ is the circular velocity at $r_{apo}$ and $\epsilon$ is the orbital eccentricity.

These are used to calculate the initial mass $M_i$ using Equation~7 of \citet{Lamers2005}:
\begin{equation}
    M_i \approx \frac{1}{\mu(t)} \times \left(\frac{M}{M_{\odot}} + \left(\frac{\gamma t}{t_0}\right)^{\frac{1}{\gamma}}\right)
\end{equation}
where $\gamma$ is the index describing mass loss, and is set to 0.7 for clusters \citep{Lamers2010}.

We use the current mass $10^{4.1}$\msun (as determined by \citealt{Simpson2018}) and the orbital parameters described in Section~\ref{sec:kinematics}.
Through this method, we find an initial mass of $10^{5.4} M_{\odot}$.

\vspace{0.4cm}
\emph{Method 2:}
We also calculate the initial mass, assuming mass loss driven by internal dynamical interactions and tidal stripping, as quantified in \citet{Baumgardt2019}.
The current mass of a cluster can be calculated by:
\begin{equation}
    M(t) = \frac{M_{\text{ini}}}{2} \left(1-\frac{t}{T_{\text{diss}}}\right)
\end{equation}
where $M_{\text{ini}}$ is the cluster's initial mass and $T_{\text{diss}}$ is the cluster's dissolution time in Myr.
The latter can be calculated with Equation~5 of \citet{Baumgardt2019}:
\begin{equation}
\label{eqn:diss_time}
T_{\text{diss}} = 1.35 \times \left(\frac{M_{\text{ini}}}{\ln (0.02N_{\text{ini}})} \right)^{0.75} \times \frac{r_{\text{apo}}}{V_c} \times (1-\epsilon)
\end{equation}
where $N_{\text{ini}}$, is the initial number of stars, which is estimated by $N_{\text{ini}} = M_{\text{ini}} / 0.65$ with the assumption that the average star's mass is 0.65 M$_{\odot}$, using the same \citet{Kroupa2001} IMF as in \citet{Baumgardt2019}.
We use the same parameters as in Method 1, as seen in Table~\ref{table:cluster_obs}.
For this method, we additionally calculate the initial mass given a range of disruption times, 11$-$14 Gyr, as we do not know exactly when the cluster was accreted onto the Milky Way.

We set the current mass equal to the mass estimate from \citet{Simpson2019}, $10^{4.1}$\msun, and solved for an initial mass which could disrupt to its current mass.
We find the initial mass is between $10^{5.5-5.7}$\msun. 
The uncertainties in mass loss are large enough that this mass estimate is relatively consistent with the estimate from the previous method.
We therefore estimate that the initial mass is in the range $10^{5.4-5.7}$\msun.
ESO~280-SC06 has therefore lost between 95 and 98\% of its initial mass. 

\section{Discussion} \label{sec:discussion}
ESO~280-SC06 is a distinctly low-metallicity and low-mass globular cluster.
By spectroscopically identifying multiple populations in this system, we can probe one of the lowest-mass gravitationally bound systems to improve our understanding of the relationship between the mass of a cluster and the chemical enrichment found therein.

\begin{figure*} 
\centering
\includegraphics[scale = 0.43]{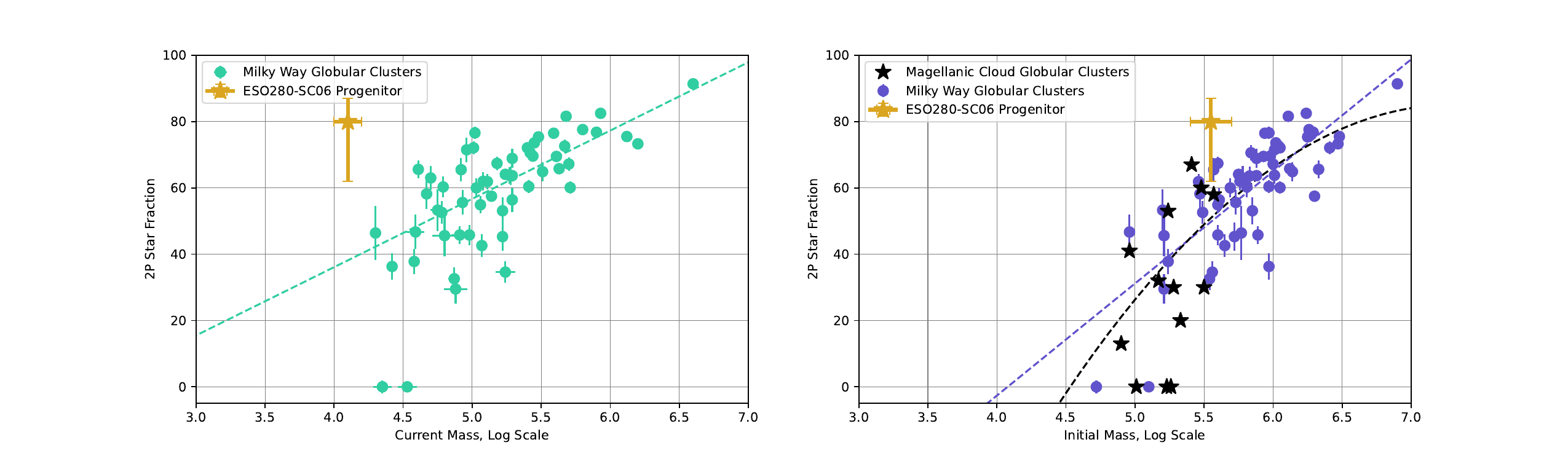}
\caption{The fraction of 2P stars in GCs as a function of current mass (left) and initial mass (right).
The current and initial masses of Milky Way globular clusters (\textit{MW GCs}) are represented by green circles on the left and purple circles on the right, respectively \citep{Milone2017, Baumgardt2019}.
Globular clusters in the Magellanic Clouds (\textit{MC GCs}), which should not experience significant mass loss, are represented by black stars.
(Data for MC GCs are from \citealt{Mucciarelli2009, Mateluna2012, Mucciarelli2014, Hollyhead2017, Niederhofer2017a, Niederhofer2017b, Martocchia2017, Hollyhead2018, Zhang2018, Martocchia2018}, and \citealt{Hollyhead2019} and were compiled by \citealt{Gratton2019}.)
ESO~280-SC06 is represented by the dark gold star.
Measurements for the 300S stellar stream are represented by a red cross \citep{Usman2024}.
The dashed and dotted lines represent linear and quadratic fits between the 2P star fraction and cluster mass.}
\label{fig:enrich_frac}
\end{figure*} 

\subsection{Enrichment Fraction vs.~Cluster Mass}\label{subsec:multipop}
The current mass of ESO~280-SC06 is $10^{4.1\pm0.1}$\msun \citep{Simpson2018}.
When compared to gravitationally bound clusters of comparable mass, the cluster appears to have a significantly high fraction of multiple population stars.
Clusters of similar current mass generally have a 2P star fraction of 40\%, as can be observed in the left panel of Figure~\ref{fig:enrich_frac}.

Given the primordial nature of GC abundance anticorrelations, one might expect that 
initial mass is a better predictor for the fraction of 2P stars \citep[e.g., ][]{Bas_Lard, Usman2024}.
When comparing our fraction of 2P stars relative to our estimated initial mass of $10^{5.55\pm0.15}$, our enrichment of 80\% still appears high relative to comparable systems, but is within error bars of the general trend.
This relationship is demonstrated in the right panel of Figure~\ref{fig:enrich_frac}.

It is possible that our sample of ten stars has an unusually high fraction of 2P stars and is not an accurate representation of the cluster.
However, we 
do not anticipate any selection effects that favor 2P stars over 1P stars. We
chose the brightest stars available to follow up in our sample, and there 
is not any known
correlation between 1P / 2P status and the brightness of the star \citep{Bas_Lard}.
We must also consider that the high fraction of 2P stars relative to initial mass could indicate that our initial mass estimate is too low. This would indicate that current models for mass loss may not fully encapsulate losses in extreme cases like ESO~280-SC06, for which the orbit never strays farther than 14 kpc from the Galactic center and crosses the plane of the disk relatively frequently.

Alternatively, ESO~280-SC06's high enrichment fraction may indicate preferential stripping of 1P stars.
\citet{Vesperini2021} carried out simulations of the impact of stellar dynamics in clusters on the spatial distribution of the 2P stars.
The authors find that clusters will preferentially have 2P stars towards the center, and 1P stars are more likely to be stripped away by external influences. 
In these simulations, the clusters will dynamically mix over time, causing a higher proportion of 2P stars to be lost at later times.
If ESO~280-SC06 experienced significant mass loss very early in its lifetime, it may have preferentially lost a higher proportion of 1P stars 
because that dynamical mixing had not yet occurred. 
Observing and chemically tagging stars that have been tidally stripped from ESO~280-SC06 could confirm or contradict this theory, because a much higher fraction of 1P stars would be expected in the population of tidally stripped stars.
Other clusters with low present-day masses may have more time to spatially homogenize prior to losing a significant portion of their stars, causing them to have an overall lower fraction of 2P stars observed today relative to ESO~280-SC06.

\subsection{NEMP Star}\label{subsec:CN_star}
In the previous work of \citet{Simpson2019}, a nitrogen-enriched metal poor star (NEMP) was identified using medium-resolution spectroscopy on 3.9-m Anglo-Australian Telescope using its AAOmega spectrograph.
We re-observe this star using high-resolution spectroscopy using Magellan/MIKE.
We find, in addition to nitrogen enhancement, the star is enhanced in \textit{s}-process elements including strontium, yttrium, barium and lanthanum.
Stars with this type of abundance anomaly are sometimes referred to as CEMP-s stars, CH stars or Ba stars, depending on the context (e.g. \citealt{Luck1991, DOrazi2010, Cote1997}).

This particular combination of chemical enrichment is indicative of mass transfer from a binary companion.
These stars are unusual in GCs, e.g. a previous study by \citet{DOrazi2010} of 1,205 red giant branch stars in GCs found just five stars of this type.
Of the five post-mass transfer stars identified in that study, just one was classified as a 2P star.
The authors suggest that high-density environments which give rise to 2P globular cluster stars are significantly more likely to disrupt binary star systems before any type of mass transfer can occur. Dynamical studies of stellar multiplicity in cluster environments \citep[e.g., ][]{Ferraro2023} find correlations between cluster density and binary evolution outcomes. 
Therefore, the presence of these \textit{s-}process enhanced stars could potentially be used as a probe of dynamical interactions within a proto-globular cluster environment.

While dense GC environments may disrupt the long-term survival of binary systems, it remains unclear whether a disrupted globular cluster would provide a more suitable environment for the survival of binary systems.
A similar post-mass transfer star was also identified in the 300S globular cluster stellar stream \citep{Usman2024}.
The detection of a post-mass transfer star in both a low-mass cluster that has undergone significant mass loss and the fully disrupted globular cluster stellar stream 300S could indicate that the same mechanism that aids the creation of these mass-transfer stars also impacts the cluster's ability to retain stars.
After all, the rate of mass-transfer stars among the globular cluster stars in \citet{DOrazi2010} is 0.4\% and in field stars is 2\% \citep{Luck1991}.
Even if star 026 is the \textit{only} mass-transfer star out of the 45 members of ESO 280-SC06 identified in \citet{Simpson2019}, ESO 280 would still have a rate of mass-transfer stars five times higher than \citet{DOrazi2010} found in GCs and comparable to the rate in the field.
Perhaps this increased rate of post-mass transfer stars hints that the weakening of the gravitational potential of a cluster both inhibits the destruction of binary star formation, and enhances the mass loss rate and subsequent dissolution of the cluster. 
Further exploration into the dynamics of binary systems in GCs during periods of significant mass loss is planned as future work as these systems could shed light on preferential natal environments for mass-transfer binaries.

\subsection{Neutron-Capture Spread}\label{subsec:ncap}
In addition to light-element variations, some GCs have variations in neutron-capture elements.
These variation patterns were first identified in M15 by \citet{Sneden1997}, which found correlated abundances of barium and europium.
These neutron-capture variations were later confirmed by \citet{Sneden2000, Otsuki2006, Sobeck2011, Worley2013} and \citet{Cabrera2024}.
The correlation of barium and europium indicates that these elements were created through an \textit{r}-process nucleosynthetic event (or events), instead of by the \textit{s}-process.
This substantially complicates the chemical evolution of GCs, because while both light element and \textit{s}-process element variations can be created in AGB stars, they are not expected to create significant amounts of \textit{r}-process elements.
Thus, additional nucleosynthetic processes are required to account for the \textit{r}-process enrichment.
Additionally, the \textit{r}-process elements in M15 have no apparent correlation with light-element abundances, i.e. there is no relationship between \textit{r}-process enrichment and 1P or 2P populations.
The lack of relationship between \textit{r}-process and multiple stellar populations indicates that these processes did not develop concurrently in M15.

Similarly, the globular cluster M92 was found to have a significant spread in \textit{r}-process elements such as yttrium, zirconium, lanthanum and europium, which again were uncorrelated with the light element enrichment patterns indicative of multiple stellar populations in \citet{Roederer2011}.
The authors found evidence of similar \textit{r}-process dispersions in M5 and NGC 3201.
In her study of 12 red giant branch stars in M92 using high-resolution spectra from Keck, \citet{Cohen2011} is unable to replicate a spread in \textit{r}-process elements.
The author demonstrates that the correlation between absorption line strength and the stellar parameters of the observed stars, and suggests that the previously identified \textit{r}-process dispersion was a result of previously unidentified systematic uncertainties.
\citet{Roederer2015} followed up on these results by analyzing 15 red giant branch stars in NGC 4833 using Magellan/MIKE and found similar heavy element spreads that could be attributed to systematics.
The authors concluded the metallicity dispersions in \citet{Roederer2011}, with the exception of M15, were likely due to these systematic uncertainties.

More recently, \citet{Kirby2023} found the \textit{r}-process spread in M92 to have an interesting relationship with the cluster's multiple stellar populations: 1P stars exhibited a strong dispersion in \textit{r}-process enrichment, while 2P exhibited very little spread.
For example, the authors find that barium, lanthanum, and europium have dispersions of $0.09^{+0.06}_{-0.05}$, $0.18^{+0.10}_{-0.06}$, and $0.14^{+0.08}_{-0.06}$ in M92's 1P stars, respectively.
In the 2P stars, however, they constrain the dispersions to $< 0.02$, $< 0.08$, and $< 0.03$, respectively
The authors suggest that an \textit{r}-process nucleosynthetic event occurred as the first stars began to form, inhomogeneously enriching the gas which formed into 1P stars and resulting in these early stars.
By the time the 2P stars formed, the star-forming gas was sufficiently mixed such that the later generation of stars had consistent \textit{r}-process enrichment.
Both of these clusters with confirmed \textit{r}-process enrichment are relatively low-metallicity, with measured [Fe/H] $\simeq -2.4$.
\citet{Kirby2023} suggested that, at higher metallicities, variations in \textit{r}-process enrichment would look insignificant relative to high levels of inherent iron enrichment in the cluster's natal gas.

We expect ESO 280-SC06 
to be a good probe of \textit{r}-process dispersion, as it is the lowest-metallicity intact globular cluster in the Milky Way at our measured metallicity of [Fe/H] $\simeq -2.54$.
The initial observations of ESO 280-SC06 presented here were intended to measure the dispersion of neutron-capture elements.
However, we are unable to measure a spread in any of our measured neutron-capture elements.
Excluding 
stars with upper limit measurements and the NEMP-s star, which is enriched in \textit{s}-process, we constrain barium, lanthanum and europium dispersions to $< 0.12$, $< 0.31$ and $< 0.18$, respectively at 95\% confidence, as shown in Table~\ref{tab:cluster_abunds}.
We may be unable to constrain these dispersions as strongly as \citet{Kirby2023} due to our conservative abundance uncertainties, which include propagated uncertainties due to stellar parameter uncertainties.
Alternatively, this may simply mean that any \textit{r}-process nucleosynthetic process in ESO 280-SC06 occurred significantly earlier than the cluster's period of star formation, allowing for sufficient gas mixing 
to create a relatively uniform enrichment pattern within the cluster.

Lastly, the lack of spread may be due to ESO~280-SC06's high 2P star fraction.
\citet{Kirby2023} specifically identifies abundance dispersions in M92's 1P stars.
Since our sample contains just two 1P stars, we may not have a high enough fraction of 1P stars to measure similar abundance spreads.
Identifying more 1P stars in ESO 280-SC06, or 1P stars that were stripped from the cluster during its orbit, may allow for a stronger constraint on neutron-capture abundance spreads.

\section{Conclusion} \label{sec:conclusion}
In this paper, we have constrained the fraction of 2P stars in the low-mass, metal-poor globular cluster ESO~280-SC06.
We find it to be surprisingly enriched, with a 2P star fraction of $\sim80\%$.
GCs of comparable current mass generally have a 2P star fraction of 40\% or less.

However, recent work suggests initial mass is a more accurate predictor of 2P star fraction than current mass \citep{Gratton2019, Usman2024}.
We use two different methods to calculate the initial mass, and conclude that ESO~280-SC06 formed with a mass in the range of $10^{5.4-5.7}$\msun.
The cluster has therefore lost approximately 95 to 98\% of its initial mass, significantly higher than the typical cluster.
In N-body simulations over various orbits, globular clusters lose on average ~80\% of their initial mass \citep{Baumgardt2019}.

Even when comparing the initial mass of the clusters, ESO~280-SC06 still appears more enriched in 2P stars relative to other Milky Way clusters.
While this could be explained by a potentially unrepresentative sample, overenrichment could indicate that current mass loss models cannot fully describe the disruption caused along extreme orbits.
ESO~280-SC06's orbit varies between 1 kpc and 14 kpc from the Galactic center, and crosses the disk 
every $\sim 150$ Myr.
This 
eccentric and tight orbit may experience stronger interactions and disruption than is currently modeled with simplified mass loss estimates.

Another possibility for the high enrichment fraction could be the preferential stripping of 1P stars in the cluster.
2P stars in GCs tend to be more centrally located than 1P stars at the time of formation, then mix over their lifetimes.
If ESO~280-SC06 experienced significant mass loss early in its lifetime, it may have not been dynamically mixed prior to experiencing tidal stripping.
This uneven spatial distribution would result in 1P stars being stripped first from the cluster, 
leaving 
a higher fraction of 2P stars relative to clusters that were fully mixed prior to experiencing significant tidal disruption.

We re-observe a nitrogen-enriched star initially identified in \citet{Simpson2019}.
We identify significant enrichment in \textit{s}-process elements, and therefore classify it as a post-mass transfer star.
Sometimes called CEMP-s stars, CH stars or Ba stars, post-mass transfer stars are not usually identified in GCs, as they require 
close binary systems that long enough for a $3-8$ \msun star to evolve onto the AGB, 
and we expect these to be disrupted in dense cluster environments.
A similar post-mass transfer star was identified in the 300S globular cluster stellar stream by \citet{Usman2024}.
This may indicate that disrupting or disrupted system may provide a more hospitable environment for binary systems than dense, stable GCs.

Lastly, we find no detectable spread in neutron-capture elements in ESO~280-SC06.
Previous studies of metal-poor GCs M15 and M92 found detectable spreads of neutron-capture elements among the 1P stars \citep{Sobeck2011, Kirby2023}.
It has been suggested that metal-poor clusters may more clearly demonstrate these dispersions than their metal-rich counterparts due to enrichment mechanisms which increase enrichment in both metallicity and neutron-capture elements.
ESO~280-SC06 may contradict this trend; however, our conservative abundance uncertainties may simply not allow for dispersions to be constrained as tightly as they have been in analyses such as \citet{Kirby2023}.
Further study could confirm ESO 280-SC06's uniform enrichment in neutron-capture elements.

\section*{Acknowledgments:}
This paper includes data gathered with the 6.5~meter Magellan Telescopes located at Las Campanas Observatory, Chile.

SAU acknowledges the support of the American Association of University Women through their American Dissertation Fellowship.

JR acknowledged the support from the Carnegie Observatory Internship and to the mentors throughout the project.

APJ acknowledges support from a Carnegie Fellowship, the Thacher Research Award in Astronomy, and the Alfred P. Sloan Research Fellowship. APJ and SAU were supported by the National Science Foundation grant AST-2206264.

SLM and JDS acknowledge support from the Australian Research Council through Discovery Project grant DP180101791 and from the UNSW Scientia Fellowship programme. This research has been supported in part by the Australian Research Council Centre of Excellence for All Sky Astrophysics in 3 Dimensions (ASTRO 3D), through project number CE170100013. SLM also acknowledges funding from ARC DP220102254.

T.S.L. acknowledge financial support from Natural Sciences and Engineering Research Council of Canada (NSERC) through grant RGPIN-2022-04794.

This research has made use of NASA’s Astrophysics Data System Bibliographic Services; the arXiv preprint server operated by Cornell University; and the SIMBAD databases hosted by the Strasbourg Astronomical Data Center.

This work has made use of data from the European Space Agency (ESA) mission
{\it Gaia} (\url{https://www.cosmos.esa.int/gaia}), processed by the {\it Gaia}
Data Processing and Analysis Consortium (DPAC,
\url{https://www.cosmos.esa.int/web/gaia/dpac/consortium}). Funding for the DPAC
has been provided by national institutions, in particular the institutions
participating in the {\it Gaia} Multilateral Agreement.

{\it Software:} 
{\code{numpy} \citep{numpy}, 
\code{matplotlib} \citep{matplotlib}, 
\code{astropy} \citep{astropy:2018},
}

\setlength\tabcolsep{0.1cm}
\begin{table}
    \centering
    \caption{Star 001 Abundances\label{tab:001}}
    \begin{tabular}{cccccccc}
Element & Species & N & $\log \epsilon$ & [X/H] & $e_{\text{[X/H]}}$ & [X/Fe] & $e_{\text{[X/Fe]}}$\\
\hline 
Fe I & 26.0 & 25 & 5.12 & $-$2.38 & 0.17 & $-$ & $-$ \\
Fe II & 26.1 & 7 & 5.02 & $-$2.48 & 0.13 & $-$ & $-$ \\
\hline
Na I & 11.0 & 2 & 4.87 & $-$1.37 & 0.18 & 1.01 & 0.2 \\
Na, nLTE &  &  &  &  &  & 0.66 &  \\
Mg I & 12.0 & 5 & 5.35 & $-$2.25 & 0.14 & 0.13 & 0.19 \\
Al I & 13.0 & 1 & 4.48 & $-$1.97 & 0.31 & 0.41 & 0.34 \\
Al, nLTE &  &  &  &  &  & 1.24 &  \\
Si I & 14.0 & 2 & 5.91 & $-$1.6 & 0.24 & 0.78 & 0.26 \\
K I & 19.0 & 2 & 3.23 & $-$1.8 & 0.2 & 0.58 & 0.23 \\
Ca I & 20.0 & 7 & 4.26 & $-$2.08 & 0.14 & 0.29 & 0.18 \\
Sc II & 21.1 & 6 & 0.86 & $-$2.29 & 0.17 & 0.19 & 0.2 \\
Ti I & 22.0 & 6 & 2.85 & $-$2.1 & 0.15 & 0.28 & 0.19 \\
Ti II & 22.1 & 13 & 2.89 & $-$2.06 & 0.18 & 0.42 & 0.2 \\
V II & 23.1 & 1 & 1.62 & $-$2.31 & 0.28 & 0.17 & 0.31 \\
Cr I & 24.0 & 1 & 3.2 & $-$2.44 & 0.19 & $-$0.07 & 0.23 \\
Mn I & 25.0 & 1 & 2.61 & $-$2.82 & 0.26 & $-$0.44 & 0.29 \\
Co I & 27.0 & 4 & 2.85 & $-$2.14 & 0.17 & 0.24 & 0.22 \\
Ni I & 28.0 & 2 & 4.04 & $-$2.18 & 0.18 & 0.2 & 0.21 \\
Zn I & 30.0 & 1 & 2.42 & $-$2.14 & 0.24 & 0.23 & 0.28 \\
Sr II & 38.1 & 2 & 0.75 & $-$2.12 & 0.25 & 0.36 & 0.26 \\
Y II & 39.1 & 3 & $-$0.29 & $-$2.5 & 0.2 & $-$0.02 & 0.23 \\
Zr II & 40.1 & 1 & 0.35 & $-$2.23 & 0.26 & 0.25 & 0.28 \\
Ba II & 56.1 & 5 & $-$0.71 & $-$2.89 & 0.17 & $-$0.42 & 0.2 \\
Eu II & 63.1 & 2 & $-$1.35 & $-$1.87 & 0.16 & 0.61 & 0.2 \\
C$-$H & 106.0 & 2 & 6.15 & $-$2.28 & 0.22 & 0.09 & 0.22 \\
C$-$N & 607.0 & 1 & 7.03 & $-$0.8 & 0.27 & 1.58 & 0.26 \\
\hline
O I & 8.0 & 1 & $-$ & $-$0.64 & $-$ & 1.73 & $-$ \\
V I & 23.0 & 1 & $-$ & $-$2.05 & $-$ & 0.33 & $-$ \\
La II & 57.1 & 4 & $-$ & $-$1.16 & $-$ & 1.32 & $-$ \\
Dy II & 66.1 & 3 & $-$ & $-$0.62 & $-$ & 1.86 & $-$ 
    \end{tabular}
\end{table}

\begin{table}
    \centering
    \caption{Star 003 Abundances}
    \begin{tabular}{cccccccc}
Element & Species & N & $\log \epsilon$ & [X/H] & $e_{\text{[X/H]}}$ & [X/Fe] & $e_{\text{[X/Fe]}}$\\
\hline 
Fe I & 26.0 & 41 & 4.9 & $-$2.6 & 0.21 & $-$ & $-$ \\
Fe II & 26.1 & 12 & 4.84 & $-$2.66 & 0.13 & $-$ & $-$ \\
\hline
Na I & 11.0 & 2 & 4.61 & $-$1.63 & 0.21 & 0.97 & 0.26 \\
Na, nLTE &  &  &  &  &  & 0.38 &   \\
Mg I & 12.0 & 4 & 5.27 & $-$2.33 & 0.14 & 0.27 & 0.23 \\
Al I & 13.0 & 1 & 4.31 & $-$2.14 & 0.27 & 0.46 & 0.31 \\
Al, nLTE &  &  &  &  &  & 1.17 &  \\
Si I & 14.0 & 2 & 5.5 & $-$2.01 & 0.25 & 0.59 & 0.3 \\
K I & 19.0 & 2 & 3.16 & $-$1.87 & 0.14 & 0.73 & 0.23 \\
Ca I & 20.0 & 14 & 4.17 & $-$2.17 & 0.14 & 0.43 & 0.23 \\
Sc II & 21.1 & 6 & 0.61 & $-$2.54 & 0.16 & 0.11 & 0.19 \\
Ti I & 22.0 & 6 & 2.67 & $-$2.28 & 0.24 & 0.31 & 0.29 \\
Ti II & 22.1 & 17 & 2.7 & $-$2.25 & 0.19 & 0.41 & 0.22 \\
Cr I & 24.0 & 1 & 3.0 & $-$2.64 & 0.18 & $-$0.04 & 0.25 \\
Cr II & 24.1 & 1 & 3.16 & $-$2.48 & 0.18 & 0.18 & 0.21 \\
Mn I & 25.0 & 1 & 2.71 & $-$2.72 & 0.3 & $-$0.12 & 0.35 \\
Co I & 27.0 & 4 & 2.59 & $-$2.4 & 0.26 & 0.2 & 0.3 \\
Ni I & 28.0 & 2 & 3.65 & $-$2.57 & 0.17 & 0.03 & 0.25 \\
Sr II & 38.1 & 2 & 0.58 & $-$2.29 & 0.32 & 0.37 & 0.33 \\
Ba II & 56.1 & 5 & $-$0.78 & $-$2.96 & 0.15 & $-$0.3 & 0.19 \\
Eu II & 63.1 & 1 & $-$1.72 & $-$2.24 & 0.24 & 0.42 & 0.27 \\
C$-$H & 106.0 & 2 & 6.13 & $-$2.3 & 0.21 & 0.3 & 0.25 \\
C$-$N & 607.0 & 1 & 6.5 & $-$1.33 & 0.35 & 1.27 & 0.36 \\
\hline
O I & 8.0 & 1 & $-$ & $-$0.99 & $-$ & 1.61 & $-$ \\
V I & 23.0 & 1 & $-$ & $-$2.15 & $-$ & 0.45 & $-$ \\
V II & 23.1 & 2 & $-$ & $-$1.47 & $-$ & 1.19 & $-$ \\
Zn I & 30.0 & 1 & $-$ & $-$2.26 & $-$ & 0.34 & $-$ \\
Y II & 39.1 & 2 & $-$ & $-$2.17 & $-$ & 0.49 & $-$ \\
Zr II & 40.1 & 1 & $-$ & $-$1.95 & $-$ & 0.71 & $-$ \\
La II & 57.1 & 3 & $-$ & $-$1.47 & $-$ & 1.18 & $-$ \\
Dy II & 66.1 & 2 & $-$ & $-$1.11 & $-$ & 1.55 & $-$ 
   \end{tabular}
   \label{tab:003}
\end{table}

\begin{table}
    \centering
    \caption{Star 005 Abundances}
    \begin{tabular}{cccccccc}
Element & Species & N & $\log \epsilon$ & [X/H] & $e_{\text{[X/H]}}$ & [X/Fe] & $e_{\text{[X/Fe]}}$\\
\hline
Fe I & 26.0 & 57 & 4.97 & $-$2.53 & 0.2 & $-$ & $-$ \\
Fe II & 26.1 & 16 & 5.03 & $-$2.47 & 0.17 & $-$ & $-$ \\
\hline
Na I & 11.0 & 2 & 4.3 & $-$1.94 & 0.19 & 0.58 & 0.23 \\
Na, nLTE &  &  &  &  &  & 0.03 &  \\
Mg I & 12.0 & 5 & 5.4 & $-$2.2 & 0.15 & 0.33 & 0.22 \\
Al I & 13.0 & 1 & 2.97 & $-$3.48 & 0.35 & $-$0.95 & 0.39 \\
Al, nLTE &  &  &  &  &  & $-$0.23 &  \\
Si I & 14.0 & 2 & 5.5 & $-$2.01 & 0.26 & 0.51 & 0.31 \\
K I & 19.0 & 2 & 3.11 & $-$1.92 & 0.14 & 0.6 & 0.22 \\
Ca I & 20.0 & 11 & 4.19 & $-$2.15 & 0.12 & 0.38 & 0.21 \\
Sc II & 21.1 & 5 & 0.69 & $-$2.46 & 0.14 & 0.01 & 0.21 \\
Ti I & 22.0 & 6 & 2.74 & $-$2.21 & 0.23 & 0.31 & 0.27 \\
Ti II & 22.1 & 20 & 2.83 & $-$2.12 & 0.16 & 0.36 & 0.21 \\
V II & 23.1 & 1 & 1.62 & $-$2.31 & 0.2 & 0.16 & 0.26 \\
Cr I & 24.0 & 1 & 3.03 & $-$2.61 & 0.18 & $-$0.09 & 0.24 \\
Cr II & 24.1 & 2 & 3.16 & $-$2.48 & 0.15 & $-$0.0 & 0.22 \\
Mn I & 25.0 & 2 & 2.53 & $-$2.9 & 0.19 & $-$0.37 & 0.26 \\
Co I & 27.0 & 4 & 2.62 & $-$2.37 & 0.17 & 0.15 & 0.24 \\
Ni I & 28.0 & 2 & 3.81 & $-$2.41 & 0.18 & 0.12 & 0.24 \\
Sr II & 38.1 & 2 & 0.76 & $-$2.11 & 0.25 & 0.36 & 0.28 \\
Y II & 39.1 & 2 & $-$0.23 & $-$2.44 & 0.23 & 0.04 & 0.28 \\
Zr II & 40.1 & 1 & 0.63 & $-$1.95 & 0.34 & 0.52 & 0.37 \\
Ba II & 56.1 & 4 & $-$0.65 & $-$2.83 & 0.15 & $-$0.35 & 0.21 \\
La II & 57.1 & 2 & $-$0.97 & $-$2.07 & 0.22 & 0.4 & 0.27 \\
\hline
O I & 8.0 & 1 & $-$ & $-$1.08 & $-$ & 1.45 & $-$ \\
V I & 23.0 & 1 & $-$ & $-$2.23 & $-$ & 0.29 & $-$ \\
Zn I & 30.0 & 1 & $-$ & $-$1.93 & $-$ & 0.6 & $-$ \\
Eu II & 63.1 & 3 & $-$ & $-$0.89 & $-$ & 1.59 & $-$ \\
Dy II & 66.1 & 3 & $-$ & $-$1.09 & $-$ & 1.38 & $-$ \\
C$-$H & 106.0 & 2 & $-$ & $-$2.4 & $-$ & 0.13 & $-$ \\
C$-$N & 607.0 & 1 & $-$ & $-$0.31 & $-$ & 2.21 & $-$
    \end{tabular}
    \label{tab:005}
\end{table}

\begin{table}
    \centering
    \caption{Star 008 Abundances}
    \begin{tabular}{cccccccc}
Element & Species & N & $\log \epsilon$ & [X/H] & $e_{\text{[X/H]}}$ & [X/Fe] & $e_{\text{[X/Fe]}}$\\
\hline
Fe I & 26.0 & 44 & 4.9 & $-$2.6 & 0.18 & $-$ & $-$ \\
Fe II & 26.1 & 13 & 5.02 & $-$2.48 & 0.17 & $-$ & $-$ \\
\hline
Na I & 11.0 & 2 & 4.43 & $-$1.81 & 0.19 & 0.79 & 0.22 \\
Na, nLTE &  &  &  &  &  & 0.23 &  \\
Mg I & 12.0 & 5 & 5.31 & $-$2.29 & 0.18 & 0.31 & 0.23 \\
Al I & 13.0 & 1 & 3.4 & $-$3.05 & 0.38 & $-$0.46 & 0.41 \\
Al, nLTE &  &  &  &  &  & 0.54 &  \\
Si I & 14.0 & 2 & 5.54 & $-$1.97 & 0.26 & 0.63 & 0.28 \\
K I & 19.0 & 2 & 3.11 & $-$1.92 & 0.14 & 0.67 & 0.2 \\
Ca I & 20.0 & 14 & 4.13 & $-$2.21 & 0.13 & 0.38 & 0.19 \\
Sc II & 21.1 & 7 & 0.67 & $-$2.48 & 0.18 & 0.01 & 0.23 \\
Ti I & 22.0 & 6 & 2.74 & $-$2.21 & 0.17 & 0.38 & 0.21 \\
Ti II & 22.1 & 25 & 2.76 & $-$2.19 & 0.18 & 0.29 & 0.22 \\
V I & 23.0 & 1 & 1.58 & $-$2.35 & 0.21 & 0.25 & 0.25 \\
V II & 23.1 & 1 & 1.52 & $-$2.41 & 0.39 & 0.08 & 0.42 \\
Cr I & 24.0 & 1 & 2.88 & $-$2.76 & 0.18 & $-$0.16 & 0.23 \\
Cr II & 24.1 & 1 & 2.98 & $-$2.66 & 0.18 & $-$0.18 & 0.23 \\
Mn I & 25.0 & 3 & 2.47 & $-$2.96 & 0.18 & $-$0.36 & 0.24 \\
Co I & 27.0 & 4 & 2.39 & $-$2.6 & 0.29 & $-$0.0 & 0.33 \\
Ni I & 28.0 & 2 & 3.52 & $-$2.7 & 0.18 & $-$0.1 & 0.23 \\
Zn I & 30.0 & 1 & 2.08 & $-$2.47 & 0.22 & 0.12 & 0.27 \\
Sr II & 38.1 & 2 & 0.5 & $-$2.37 & 0.29 & 0.11 & 0.31 \\
Y II & 39.1 & 1 & $-$0.44 & $-$2.65 & 0.21 & $-$0.17 & 0.26 \\
Zr II & 40.1 & 1 & 0.42 & $-$2.16 & 0.27 & 0.32 & 0.31 \\
Ba II & 56.1 & 4 & $-$0.57 & $-$2.75 & 0.15 & $-$0.27 & 0.2 \\
La II & 57.1 & 1 & $-$1.35 & $-$2.45 & 0.28 & 0.03 & 0.32 \\
Eu II & 63.1 & 2 & $-$1.83 & $-$2.35 & 0.19 & 0.13 & 0.24 \\
\hline
O I & 8.0 & 1 & $-$ & $-$1.17 & $-$ & 1.42 & $-$ \\
Dy II & 66.1 & 2 & $-$ & $-$1.19 & $-$ & 1.29 & $-$ \\
C$-$H & 106.0 & 2 & $-$ & $-$2.76 & $-$ & $-$0.16 & $-$ \\
C$-$N & 607.0 & 1 & $-$ & $-$1.0 & $-$ & 1.6 & $-$ 
    \end{tabular}
    \label{tab:008}
\end{table}

\begin{table}
    \centering
    \caption{Star 016 Abundances}
    \begin{tabular}{cccccccc}
Element & Species & N & $\log \epsilon$ & [X/H] & $e_{\text{[X/H]}}$ & [X/Fe] & $e_{\text{[X/Fe]}}$\\
\hline
Fe I & 26.0 & 43 & 5.06 & $-$2.44 & 0.15 & $-$ & $-$ \\
Fe II & 26.1 & 11 & 4.98 & $-$2.52 & 0.13 & $-$ & $-$ \\
\hline
Na I & 11.0 & 2 & 4.69 & $-$1.55 & 0.19 & 0.89 & 0.19 \\
Na, nLTE &  &  &  &  &  & 0.40 &  \\
Mg I & 12.0 & 6 & 5.56 & $-$2.04 & 0.14 & 0.4 & 0.17 \\
Al I & 13.0 & 1 & 4.48 & $-$1.97 & 0.21 & 0.47 & 0.24 \\
Al, nLTE &  &  &  &  &  & 1.23 &  \\
Si I & 14.0 & 1 & 5.64 & $-$1.87 & 0.24 & 0.57 & 0.25 \\
K I & 19.0 & 2 & 3.32 & $-$1.71 & 0.15 & 0.73 & 0.18 \\
Ca I & 20.0 & 16 & 4.21 & $-$2.13 & 0.16 & 0.31 & 0.19 \\
Sc II & 21.1 & 7 & 0.84 & $-$2.31 & 0.14 & 0.2 & 0.16 \\
Ti I & 22.0 & 10 & 2.77 & $-$2.18 & 0.22 & 0.25 & 0.23 \\
Ti II & 22.1 & 27 & 2.85 & $-$2.1 & 0.17 & 0.41 & 0.18 \\
V I & 23.0 & 1 & 1.46 & $-$2.47 & 0.31 & $-$0.03 & 0.33 \\
V II & 23.1 & 2 & 1.49 & $-$2.44 & 0.19 & 0.07 & 0.22 \\
Cr I & 24.0 & 1 & 2.76 & $-$2.88 & 0.2 & $-$0.44 & 0.22 \\
Cr II & 24.1 & 1 & 2.93 & $-$2.71 & 0.2 & $-$0.19 & 0.23 \\
Mn I & 25.0 & 4 & 2.53 & $-$2.9 & 0.15 & $-$0.46 & 0.19 \\
Co I & 27.0 & 5 & 2.77 & $-$2.22 & 0.19 & 0.22 & 0.21 \\
Ni I & 28.0 & 4 & 3.79 & $-$2.43 & 0.16 & 0.01 & 0.18 \\
Zn I & 30.0 & 1 & 2.19 & $-$2.36 & 0.34 & 0.07 & 0.36 \\
Sr II & 38.1 & 2 & 0.7 & $-$2.17 & 0.21 & 0.35 & 0.21 \\
Y II & 39.1 & 4 & $-$0.46 & $-$2.67 & 0.15 & $-$0.15 & 0.18 \\
Zr II & 40.1 & 1 & 0.17 & $-$2.41 & 0.2 & 0.11 & 0.23 \\
Ba II & 56.1 & 5 & $-$0.54 & $-$2.72 & 0.15 & $-$0.2 & 0.16 \\
La II & 57.1 & 1 & $-$1.64 & $-$2.74 & 0.3 & $-$0.22 & 0.32 \\
Eu II & 63.1 & 2 & $-$1.58 & $-$2.1 & 0.16 & 0.42 & 0.19 \\
Dy II & 66.1 & 2 & $-$0.78 & $-$1.88 & 0.26 & 0.64 & 0.28 \\
C$-$H & 106.0 & 2 & 5.94 & $-$2.49 & 0.22 & $-$0.05 & 0.22 \\
C$-$N & 607.0 & 1 & 6.58 & $-$1.25 & 0.31 & 1.19 & 0.29 \\
\hline
O I & 8.0 & 1 & $-$ & $-$1.1 & $-$ & 1.34 & $-$
    \end{tabular}
    \label{tab:016}
\end{table}

\begin{table}
    \centering
    \caption{Star 025 Abundances}
    \begin{tabular}{cccccccc}
Element & Species & N & $\log \epsilon$ & [X/H] & $e_{\text{[X/H]}}$ & [X/Fe] & $e_{\text{[X/Fe]}}$\\
\hline
Fe I & 26.0 & 28 & 4.9 & $-$2.6 & 0.2 & $-$ & $-$ \\
Fe II & 26.1 & 15 & 4.92 & $-$2.58 & 0.16 & $-$ & $-$ \\
\hline
Na I & 11.0 & 2 & 4.32 & $-$1.92 & 0.19 & 0.68 & 0.24 \\
Na, nLTE &  &  &  &  &  & 0.14  &  \\
Mg I & 12.0 & 5 & 5.47 & $-$2.13 & 0.14 & 0.48 & 0.22 \\
Al I & 13.0 & 1 & 2.73 & $-$3.72 & 0.36 & $-$1.12 & 0.41 \\
Al, nLTE &  &  &  &  &  & $-$0.16 &  \\
Si I & 14.0 & 2 & 5.52 & $-$1.99 & 0.19 & 0.61 & 0.25 \\
K I & 19.0 & 2 & 3.29 & $-$1.74 & 0.14 & 0.86 & 0.23 \\
Ca I & 20.0 & 11 & 4.18 & $-$2.16 & 0.12 & 0.45 & 0.21 \\
Sc II & 21.1 & 6 & 0.78 & $-$2.37 & 0.14 & 0.21 & 0.2 \\
Ti I & 22.0 & 5 & 2.73 & $-$2.22 & 0.18 & 0.38 & 0.24 \\
Ti II & 22.1 & 20 & 2.77 & $-$2.18 & 0.16 & 0.4 & 0.21 \\
V I & 23.0 & 1 & 1.66 & $-$2.27 & 0.23 & 0.33 & 0.28 \\
V II & 23.1 & 2 & 1.74 & $-$2.19 & 0.23 & 0.38 & 0.27 \\
Cr I & 24.0 & 2 & 2.98 & $-$2.66 & 0.26 & $-$0.05 & 0.31 \\
Mn I & 25.0 & 3 & 2.73 & $-$2.7 & 0.21 & $-$0.1 & 0.27 \\
Co I & 27.0 & 2 & 2.55 & $-$2.44 & 0.22 & 0.16 & 0.27 \\
Ni I & 28.0 & 2 & 3.67 & $-$2.55 & 0.16 & 0.06 & 0.23 \\
Zn I & 30.0 & 1 & 2.38 & $-$2.18 & 0.2 & 0.42 & 0.27 \\
Sr II & 38.1 & 2 & $-$0.13 & $-$3.0 & 0.28 & $-$0.43 & 0.31 \\
Y II & 39.1 & 2 & $-$0.45 & $-$2.66 & 0.23 & $-$0.09 & 0.27 \\
Zr II & 40.1 & 1 & 0.12 & $-$2.46 & 0.25 & 0.12 & 0.29 \\
Ba II & 56.1 & 4 & $-$0.85 & $-$3.03 & 0.15 & $-$0.45 & 0.21 \\
La II & 57.1 & 1 & $-$1.41 & $-$2.51 & 0.36 & 0.07 & 0.39 \\
\hline
O I & 8.0 & 1 & $-$ & $-$1.13 & $-$ & 1.47 & $-$ \\
Eu II & 63.1 & 3 & $-$ & $-$1.05 & $-$ & 1.53 & $-$ \\
Dy II & 66.1 & 3 & $-$ & $-$1.05 & $-$ & 1.52 & $-$ \\
C$-$H & 106.0 & 2 & $-$ & $-$2.51 & $-$ & 0.09 & $-$ \\
C$-$N & 607.0 & 1 & $-$ & $-$0.66 & $-$ & 1.94 & $-$
    \end{tabular}
    \label{tab:025}
\end{table}

\begin{table}
    \centering
    \caption{Star 026 Abundances}
    \begin{tabular}{cccccccc}
Element & Species & N & $\log \epsilon$ & [X/H] & $e_{\text{[X/H]}}$ & [X/Fe] & $e_{\text{[X/Fe]}}$\\
\hline
Fe I & 26.0 & 55 & 4.98 & $-$2.52 & 0.17 & $-$ & $-$ \\
Fe II & 26.1 & 14 & 5.08 & $-$2.42 & 0.18 & $-$ & $-$ \\
\hline
Na I & 11.0 & 2 & 4.85 & $-$1.39 & 0.19 & 1.13 & 0.2 \\
Na, nLTE &  &  &  & &  & 0.72   &  \\
Mg I & 12.0 & 5 & 5.47 & $-$2.13 & 0.14 & 0.39 & 0.18 \\
Al I & 13.0 & 1 & 3.66 & $-$2.79 & 0.32 & $-$0.27 & 0.34 \\
Al, nLTE &  &  &  &  &  & 0.69 &  \\
Si I & 14.0 & 2 & 5.78 & $-$1.73 & 0.46 & 0.79 & 0.47 \\
K I & 19.0 & 2 & 3.11 & $-$1.92 & 0.2 & 0.6 & 0.23 \\
Ca I & 20.0 & 14 & 4.23 & $-$2.11 & 0.13 & 0.42 & 0.18 \\
Sc II & 21.1 & 5 & 0.73 & $-$2.42 & 0.18 & $-$0.0 & 0.23 \\
Ti I & 22.0 & 7 & 2.78 & $-$2.17 & 0.21 & 0.35 & 0.23 \\
Ti II & 22.1 & 13 & 2.78 & $-$2.17 & 0.14 & 0.25 & 0.2 \\
V I & 23.0 & 1 & 1.49 & $-$2.44 & 0.2 & 0.08 & 0.26 \\
V II & 23.1 & 1 & 1.78 & $-$2.15 & 0.22 & 0.27 & 0.28 \\
Cr I & 24.0 & 1 & 2.72 & $-$2.92 & 0.19 & $-$0.4 & 0.22 \\
Cr II & 24.1 & 1 & 3.22 & $-$2.42 & 0.18 & $-$0.01 & 0.24 \\
Mn I & 25.0 & 1 & 2.5 & $-$2.93 & 0.2 & $-$0.41 & 0.24 \\
Co I & 27.0 & 3 & 2.52 & $-$2.47 & 0.21 & 0.05 & 0.23 \\
Ni I & 28.0 & 3 & 3.64 & $-$2.58 & 0.18 & $-$0.06 & 0.21 \\
Zn I & 30.0 & 1 & 2.14 & $-$2.41 & 0.2 & 0.11 & 0.25 \\
Sr II & 38.1 & 2 & 1.51 & $-$1.36 & 0.2 & 1.06 & 0.26 \\
Y II & 39.1 & 5 & 0.16 & $-$2.05 & 0.15 & 0.37 & 0.22 \\
Zr II & 40.1 & 1 & 0.61 & $-$1.97 & 0.22 & 0.45 & 0.27 \\
Ba II & 56.1 & 5 & 1.07 & $-$1.11 & 0.18 & 1.3 & 0.23 \\
La II & 57.1 & 5 & $-$0.22 & $-$1.32 & 0.15 & 1.09 & 0.22 \\
Eu II & 63.1 & 2 & $-$1.39 & $-$1.91 & 0.18 & 0.51 & 0.24 \\
C$-$H & 106.0 & 2 & 6.38 & $-$2.05 & 0.22 & 0.47 & 0.23 \\
C$-$N & 607.0 & 1 & 7.44 & $-$0.4 & 0.29 & 2.13 & 0.28 \\
\hline
O I & 8.0 & 1 & $-$ & $-$1.1 & $-$ & 1.42 & $-$ \\
Dy II & 66.1 & 1 & $-$ & $-$0.97 & $-$ & 1.44 & $-$
    \end{tabular}
    \label{tab:026}
\end{table}

\begin{table}
    \centering
    \caption{Star 033 Abundances}
    \begin{tabular}{cccccccc}
Element & Species & N & $\log \epsilon$ & [X/H] & $e_{\text{[X/H]}}$ & [X/Fe] & $e_{\text{[X/Fe]}}$\\
\hline
Fe I & 26.0 & 45 & 4.87 & $-$2.63 & 0.15 & $-$ & $-$ \\
Fe II & 26.1 & 15 & 4.93 & $-$2.57 & 0.13 & $-$ & $-$ \\
\hline
Na I & 11.0 & 2 & 4.48 & $-$1.76 & 0.21 & 0.87 & 0.2 \\
Na, nLTE &  &  &  &  &  & 0.44 &  \\
Mg I & 12.0 & 5 & 5.34 & $-$2.26 & 0.15 & 0.36 & 0.17 \\
Al I & 13.0 & 1 & 4.25 & $-$2.2 & 0.32 & 0.42 & 0.33 \\
Al, nLTE &  &  &  &  &  & 1.16 &  \\
Si I & 14.0 & 2 & 5.46 & $-$2.05 & 0.21 & 0.58 & 0.22 \\
K I & 19.0 & 2 & 3.09 & $-$1.94 & 0.16 & 0.69 & 0.19 \\
Ca I & 20.0 & 17 & 4.11 & $-$2.23 & 0.17 & 0.39 & 0.2 \\
Sc II & 21.1 & 8 & 0.6 & $-$2.55 & 0.17 & 0.02 & 0.19 \\
Ti I & 22.0 & 11 & 2.55 & $-$2.4 & 0.18 & 0.23 & 0.19 \\
Ti II & 22.1 & 28 & 2.71 & $-$2.24 & 0.17 & 0.33 & 0.19 \\
V I & 23.0 & 2 & 1.45 & $-$2.48 & 0.16 & 0.15 & 0.2 \\
V II & 23.1 & 2 & 1.65 & $-$2.28 & 0.19 & 0.29 & 0.21 \\
Cr I & 24.0 & 6 & 2.99 & $-$2.65 & 0.21 & $-$0.02 & 0.23 \\
Cr II & 24.1 & 1 & 3.08 & $-$2.56 & 0.19 & 0.01 & 0.21 \\
Mn I & 25.0 & 4 & 2.35 & $-$3.08 & 0.16 & $-$0.45 & 0.19 \\
Co I & 27.0 & 3 & 2.5 & $-$2.49 & 0.24 & 0.14 & 0.26 \\
Ni I & 28.0 & 3 & 3.88 & $-$2.34 & 0.16 & 0.29 & 0.18 \\
Zn I & 30.0 & 1 & 2.26 & $-$2.3 & 0.19 & 0.33 & 0.24 \\
Sr II & 38.1 & 2 & 0.37 & $-$2.5 & 0.29 & 0.07 & 0.31 \\
Y II & 39.1 & 5 & $-$0.48 & $-$2.69 & 0.15 & $-$0.12 & 0.18 \\
Zr II & 40.1 & 1 & 0.25 & $-$2.33 & 0.21 & 0.24 & 0.23 \\
Ba II & 56.1 & 5 & $-$0.76 & $-$2.94 & 0.14 & $-$0.37 & 0.17 \\
Eu II & 63.1 & 2 & $-$1.77 & $-$2.29 & 0.21 & 0.28 & 0.23 \\
C$-$H & 106.0 & 2 & 5.64 & $-$2.79 & 0.21 & $-$0.16 & 0.21 \\
C$-$N & 607.0 & 1 & 6.64 & $-$1.19 & 0.33 & 1.43 & 0.29 \\
\hline
O I & 8.0 & 1 & $-$ & $-$1.95 & $-$ & 0.68 & $-$ \\
La II & 57.1 & 3 & $-$ & $-$1.79 & $-$ & 0.78 & $-$ \\
Dy II & 66.1 & 1 & $-$ & $-$1.7 & $-$ & 0.87 & $-$ 
    \end{tabular}
    \label{tab:033}
\end{table}

\begin{table}
    \centering
    \caption{Star 184 Abundances\label{tab:576}}
    \begin{tabular}{cccccccc}
Element & Species & N & $\log \epsilon$ & [X/H] & $e_{\text{[X/H]}}$ & [X/Fe] & $e_{\text{[X/Fe]}}$\\
\hline
Fe I & 26.0 & 62 & 5.08 & $-$2.42 & 0.31 & $-$ & $-$ \\
Fe II & 26.1 & 21 & 4.94 & $-$2.56 & 0.14 & $-$ & $-$ \\
\hline
Na I & 11.0 & 2 & 4.22 & $-$2.02 & 0.45 & 0.4 & 0.3 \\
Na, nLTE &  &  &  &  &  & 0.15 &  \\
Mg I & 12.0 & 5 & 5.78 & $-$1.82 & 0.27 & 0.6 & 0.25 \\
Al I & 13.0 & 1 & 4.09 & $-$2.36 & 0.39 & 0.06 & 0.34 \\
Al, nLTE &  &  &  &  &  & 1.06 &  \\
Si I & 14.0 & 2 & 5.43 & $-$2.08 & 0.37 & 0.34 & 0.31 \\
K I & 19.0 & 2 & 3.5 & $-$1.53 & 0.33 & 0.89 & 0.26 \\
Ca I & 20.0 & 20 & 4.33 & $-$2.01 & 0.23 & 0.41 & 0.25 \\
Sc II & 21.1 & 8 & 0.81 & $-$2.34 & 0.19 & 0.22 & 0.22 \\
Ti I & 22.0 & 26 & 2.84 & $-$2.11 & 0.35 & 0.31 & 0.26 \\
Ti II & 22.1 & 26 & 2.99 & $-$1.96 & 0.19 & 0.59 & 0.22 \\
V I & 23.0 & 1 & 1.5 & $-$2.43 & 0.31 & $-$0.01 & 0.28 \\
V II & 23.1 & 1 & 1.69 & $-$2.24 & 0.23 & 0.32 & 0.27 \\
Cr I & 24.0 & 12 & 3.11 & $-$2.53 & 0.32 & $-$0.11 & 0.23 \\
Cr II & 24.1 & 2 & 3.2 & $-$2.44 & 0.16 & 0.12 & 0.19 \\
Mn I & 25.0 & 3 & 2.53 & $-$2.9 & 0.25 & $-$0.48 & 0.24 \\
Co I & 27.0 & 4 & 2.92 & $-$2.07 & 0.33 & 0.35 & 0.28 \\
Ni I & 28.0 & 14 & 3.82 & $-$2.4 & 0.31 & 0.02 & 0.27 \\
Zn I & 30.0 & 2 & 2.2 & $-$2.36 & 0.16 & 0.06 & 0.31 \\
Sr II & 38.1 & 2 & 0.75 & $-$2.12 & 0.22 & 0.44 & 0.26 \\
Y II & 39.1 & 7 & $-$0.4 & $-$2.61 & 0.18 & $-$0.06 & 0.22 \\
Zr II & 40.1 & 1 & 0.49 & $-$2.09 & 0.27 & 0.47 & 0.27 \\
Ba II & 56.1 & 5 & $-$0.65 & $-$2.83 & 0.17 & $-$0.27 & 0.2 \\
La II & 57.1 & 4 & $-$1.74 & $-$2.84 & 0.19 & $-$0.28 & 0.23 \\
Eu II & 63.1 & 3 & $-$1.87 & $-$2.39 & 0.17 & 0.17 & 0.22 \\
Dy II & 66.1 & 1 & $-$1.28 & $-$2.38 & 0.29 & 0.18 & 0.31 \\
C$-$H & 106.0 & 1 & 5.48 & $-$2.95 & 0.47 & $-$0.53 & 0.34 \\
C$-$N & 607.0 & 1 & 6.04 & $-$1.79 & 0.58 & 0.63 & 0.43 \\
\hline
O I & 8.0 & 1 & $-$ & $-$1.52 & $-$ & 0.9 & $-$
    \end{tabular}
    \label{tab:184}
\end{table}

\begin{table}
    \centering
    \caption{Star 576 Abundances}
    \begin{tabular}{cccccccc}
Element & Species & N & $\log \epsilon$ & [X/H] & $e_{\text{[X/H]}}$ & [X/Fe] & $e_{\text{[X/Fe]}}$\\
\hline
Fe I & 26.0 & 35 & 4.89 & $-$2.61 & 0.14 & $-$ & $-$ \\
Fe II & 26.1 & 16 & 4.82 & $-$2.68 & 0.12 & $-$ & $-$ \\
\hline
Na I & 11.0 & 2 & 4.36 & $-$1.88 & 0.24 & 0.74 & 0.2 \\
Na, nLTE &  &  &  &  &  & 0.43 &  \\
Mg I & 12.0 & 5 & 5.26 & $-$2.34 & 0.16 & 0.27 & 0.16 \\
Al I & 13.0 & 1 & 4.34 & $-$2.11 & 0.24 & 0.5 & 0.25 \\
Al, nLTE &  &  &  &  &  & 1.47 &  \\
Si I & 14.0 & 2 & 5.47 & $-$2.04 & 0.2 & 0.57 & 0.19 \\
K I & 19.0 & 2 & 3.12 & $-$1.91 & 0.16 & 0.7 & 0.16 \\
Ca I & 20.0 & 21 & 4.08 & $-$2.26 & 0.15 & 0.35 & 0.17 \\
Sc II & 21.1 & 7 & 0.49 & $-$2.66 & 0.13 & 0.02 & 0.15 \\
Ti I & 22.0 & 20 & 2.49 & $-$2.46 & 0.17 & 0.15 & 0.16 \\
Ti II & 22.1 & 34 & 2.63 & $-$2.32 & 0.18 & 0.35 & 0.2 \\
V I & 23.0 & 1 & 1.34 & $-$2.59 & 0.18 & 0.02 & 0.19 \\
V II & 23.1 & 2 & 1.5 & $-$2.43 & 0.17 & 0.24 & 0.19 \\
Cr I & 24.0 & 12 & 2.84 & $-$2.8 & 0.17 & $-$0.19 & 0.16 \\
Cr II & 24.1 & 3 & 3.04 & $-$2.6 & 0.13 & 0.08 & 0.16 \\
Mn I & 25.0 & 4 & 2.46 & $-$2.97 & 0.15 & $-$0.36 & 0.16 \\
Co I & 27.0 & 4 & 2.42 & $-$2.57 & 0.23 & 0.05 & 0.22 \\
Ni I & 28.0 & 16 & 3.71 & $-$2.51 & 0.23 & 0.1 & 0.23 \\
Zn I & 30.0 & 3 & 2.15 & $-$2.41 & 0.13 & 0.2 & 0.18 \\
Sr II & 38.1 & 2 & 0.19 & $-$2.68 & 0.25 & $-$0.01 & 0.26 \\
Y II & 39.1 & 6 & $-$0.67 & $-$2.88 & 0.13 & $-$0.21 & 0.16 \\
Zr II & 40.1 & 1 & 0.17 & $-$2.41 & 0.19 & 0.27 & 0.21 \\
Ba II & 56.1 & 5 & $-$0.94 & $-$3.12 & 0.15 & $-$0.44 & 0.17 \\
La II & 57.1 & 4 & $-$1.56 & $-$2.66 & 0.14 & 0.02 & 0.17 \\
Eu II & 63.1 & 2 & $-$1.76 & $-$2.28 & 0.15 & 0.4 & 0.18 \\
Dy II & 66.1 & 1 & $-$1.14 & $-$2.24 & 0.26 & 0.43 & 0.27 \\
C$-$H & 106.0 & 2 & 5.2 & $-$3.23 & 0.19 & $-$0.62 & 0.18 \\
C$-$N & 607.0 & 1 & 6.59 & $-$1.24 & 0.26 & 1.37 & 0.22 \\
\hline
O I & 8.0 & 1 & $-$ & $-$1.62 & $-$ & 0.99 & $-$ 
    \end{tabular}
\end{table}

\bibliography{main}{}
\bibliographystyle{aasjournal}

\end{document}